\documentstyle{article}
\def\lsim{\mathrel{\rlap{\lower 4pt \hbox{\hskip 1pt $\sim$}}\raise 1pt \hbox
	{$<$}}}
\def\gsim{\mathrel{\rlap{\lower 4pt \hbox{\hskip 1pt $\sim$}}\raise 1pt \hbox
	{$>$}}}

\newcommand\mpl{m_{{\rm Pl}}}
\newcommand\rep{\sqrt{8\pi}}
\newcommand\mps{\frac {m_{{\rm Pl}}^2}{8\pi} }
\newcommand\mpsinv{\frac{8\pi}{m_{{\rm Pl}}^2} }

\begin{document}
\begin{titlepage}
\begin{flushright}
KUNS 1207 \\
LANCASTER-TH 9401\\
SUSSEX-AST 94/1-1 \\
astro-ph/9401011 \\
(January 1994)\\
\end{flushright}
\begin{center}
\LARGE

{\bf False Vacuum Inflation with Einstein Gravity}\\
\vspace{.3in}
\normalsize
\large{Edmund J.~Copeland$^1$, Andrew R.~Liddle$^1$, David H.~Lyth$^2$,\\
\vspace*{4pt}
Ewan D.~Stewart$^3$ and David Wands$^1$} \\
\normalsize
\vspace{.6 cm}
{\em $^1$School of Mathematical and Physical Sciences, \\
University of Sussex, \\ Falmer, Brighton BN1 9QH,~~~U.~K.}\\
\vspace{.6 cm}
{\em $^2$School of Physics and Materials, \\ Lancaster University, \\
Lancaster LA1 4YB,~~~U.~K.}\\
\vspace{.6 cm}
{\em $^3$Department of Physics, \\ Kyoto University, \\
Kyoto 606,~~~Japan}
\vspace{.6 cm}
\end{center}
\baselineskip=24pt
\begin{abstract}
\noindent
We present a detailed investigation of chaotic inflation models which feature
two scalar fields, such that one field (the inflaton) rolls while the other is
trapped in a false vacuum state. The false vacuum becomes unstable when the
magnitude of the inflaton field falls below some critical value, and a first
or second order transition to the true vacuum ensues. Particular
attention is paid to the case, termed `Hybrid Inflation' by Linde, where the
false vacuum energy density dominates, so that the phase transition signals
the end of inflation. We focus mostly on the case of a second order transition,
but treat also the first order case and discuss bubble production in that
context for the first time.

False vacuum dominated inflation is dramatically different from the usual
true vacuum case, both in its cosmology and in its relation to particle
physics. The spectral index of the adiabatic density perturbation
originating during inflation can be indistinguishable from 1, or it can be up
to ten percent or so higher. The energy scale at the end of inflation
can be anywhere between $10^{16}$\,GeV, which is familiar from the true vacuum
case, and $10^{11}$\,GeV. On the other hand reheating is prompt, so the
reheat temperature cannot be far below $10^{11}\,$GeV. Cosmic strings
or other topological defects are almost inevitably produced at the end
of inflation, and if the inflationary energy scale is near its upper
limit they contribute significantly to large scale structure formation
and the cosmic microwave background anisotropy.

Turning to the particle physics, false vacuum inflation occurs with the
inflaton field far below the Planck scale and is therefore somewhat
easier to implement in the context of supergravity than true vacuum chaotic
inflation. The smallness of the inflaton mass compared with the inflationary
Hubble parameter still presents a difficulty for generic supergravity
theories. Remarkably however, the difficulty can be avoided in a natural way
for a class of supergravity models that follow from orbifold compactification
of superstrings. This opens up the prospect of a truly realistic,
superstring derived theory of inflation. One possibility, which we
show to be viable at least in the context of global supersymmetry,
is that the Peccei-Quinn symmetry is responsible for the false vacuum.
\end{abstract}

\begin{center}
{\small PACS numbers: \hspace{0.5cm} 98.80.Cq, 04.50.+h}
\end{center}
%%%%%%%%%%%%%%%%%%%%%%%%%%%%%%%%%%%%%%%%%%%%%%%%%%%%%%%%%%%%%%%%%%%%%%
\end{titlepage}
%%%%%%%%%%%%%%%%%%%%%%%%%%%%%%%%%%%%%%%%%%%%%%%%%%%%%%%%%%%%%%%%%%%%%%
%\double

\section{Introduction}
\setcounter{equation}{0}
\def\theequation{\thesection.\arabic{equation}}

An attractive proposal concerning the first moments of the
observable universe is that of chaotic inflation \cite{CHAOTIC}. At some
initial epoch, presumably the Planck scale, the various scalar fields existing
in nature are roughly homogeneous and dominate the energy density. Their
initial values are random, subject to the constraint that the energy density
is at the Planck scale. Amongst them is the inflaton field $\phi$, which is
distinguished from the non-inflaton fields by the fact that the potential is
relatively flat in its direction. Before the inflaton field $\phi$ has had
time to change much, the non-inflaton fields quickly settle down to their
minimum at fixed $\phi$, after which inflation occurs as $\phi$ rolls slowly
down the potential.

Two possibilities exist concerning the minimum into which the non-inflaton
fields fall. The simplest possibility is that it corresponds to the true
vacuum; that is, the non-inflaton fields have the same values as in the
present universe. Inflation then ends when the inflaton field starts to
execute decaying oscillations around its own vacuum value, and the hot Big
Bang (`reheating') ensues when the vacuum value has been achieved and the
decay products have thermalised. This is the usually considered case, which
has been widely explored. The other possibility is that the minimum
corresponds to a
false vacuum, with non-zero energy density. This case may be called {\it false
vacuum inflation}, and is the subject of the present paper.

There are two fundamentally different kinds of false vacuum inflation,
according to whether the energy density is dominated by the false vacuum
energy density or by the potential energy of the inflaton field. (For
simplicity we discount for the moment the intermediate possibility that the
two contributions are comparable, though it will be dealt with in the body of
the paper.) In all cases the false vacuum exists only when the value of the
inflaton field is above some critical value. If the false vacuum energy
dominates, a phase transition occurs promptly when the inflaton field falls
below the critical value, causing the end of inflation and prompt reheating.
The result is a new model of inflation which is dramatically different from
the usual one, and at least as attractive. It was first studied by Linde who
termed it `Hybrid Inflation', and it is the main focus of the present paper.
The
phase transition may be of either first or second order. A first-order model
of false vacuum dominated inflation has been considered by Linde
\cite{Linde90} and (with minor differences but more thoroughly) by Adams and
Freese \cite{AdamsFreese}. A second-order model has been discussed by Linde
\cite{LIN2SC,LIN2SC2} and explored in a preliminary way by Liddle and Lyth
\cite{LL2} and by Mollerach, Matarrese and Lucchin \cite{MML}. As far as we
know these are the only references in the literature to false vacuum dominated
inflation with Einstein gravity. Related models have been considered at some
length in the context of extended gravity theories
\cite{LaSteinhardt,extinf,Amendola}; although such theories can be recast as
Einstein gravity theories by a conformal transformation, the resulting
potentials are of a different type and this case is excluded from the present
paper.

The opposite case where the false vacuum energy is negligible (inflaton
domination) is indistiguishable from the true vacuum case for couplings of
order unity, though a variety of exotic effects can occur for small couplings.
This case has been studied by several authors
\cite{Kofman,vishniac,kofpog,yokoyama,kbhp,sbb,lyth90,hodpri,nagasawa},
and in
the present paper it is treated fairly briefly.

{}From the viewpoint of cosmology, false vacuum dominated inflation differs
from
the usual true vacuum case in three important respects.
\begin{enumerate}
\item The spectral index $n$ of the adiabatic density perturbation is
typically very close to the scale invariant value 1, and is in any case greater
than 1. This is in contrast with other working models of inflation, where one
typically finds $n<1$, viable models covering a range from perhaps
$n \simeq 0.7$ up to
$n \simeq 1$ \cite{LL2}. We shall however note that the extent to which
$n$ can exceed unity is quite limited, contrary to claims in
Refs.~\cite{LIN2SC2,MML}.
\item Topological defects generally form at the {\it end} of inflation, in
accordance with the homotopy groups of the breaking of the false vacuum to
degenerate states, provided that these groups exist. The defects may be
of any type (domain
walls, gauge or global strings, gauge or global monopoles, textures or
nontopological textures).
\item Reheating occurs promptly at the end of inflation. In the simple
models that we have explored, this means that the reheat temperature is
at least $10^{11}$\,GeV. One consequence is that a long lived
gravitino must be either rather heavy
($m\gsim 1$\,TeV)
or extremely light, so as not to be overproduced \cite{kawashi}.
\end{enumerate}

False
vacuum dominated
inflation is also very different from the true vacuum case
from the viewpoint of particle physics. Sticking to the chaotic
inflation scenario already described,
let us consider as a specific example the inflationary potential
\begin{equation}
V(\phi)=V_0+ \frac12 m^2\phi^2+\frac14\lambda \phi^4 \,.
\end{equation}
where $V_0$ is the false vacuum
energy density. Consider first the true vacuum case, where $V_0$
vanishes. Inflation occurs while $\phi$ rolls slowly towards zero, and
it ends when $\phi$ begins to oscillate, which occurs when $\phi$ is of
order the Planck mass. In order to have sufficiently small
cosmic microwave background (cmb)
anisotropy, one needs $m\lsim 10^{13}\,$GeV and $\lambda\lsim 10^{-12}$,
with one or conceivably both of these limits saturated if inflation
is to actually generate the observed anisotropy (and a primeval
density perturbation leading to structure formation).
To achieve the small $\lambda$ in a natural way one should invoke
supersymmetry. As long as one sticks to global supersymmetry this
presents no problem, but there are sound particle physics reasons
for invoking instead local supersymmetry, which is termed
supergravity because it automatically includes gravity.
In the context of supergravity, the fact that
$\phi$ is of order the Planck mass during inflation
is problematical, because in this regime
it is difficult to arrange for a sufficiently flat potential.

As will become clear, things are very different in the false vacuum case.
One still needs to have $\lambda$ very small, and will still
therefore wish to implement inflation
in the context of supergravity. But now
$\phi$ is far below the Planck scale during
inflation (after the observable universe leaves the
horizon which is the cosmologically interesting era).
As a result it becomes easier to construct  a viable model of inflation,
though the smallness of $m$ in relation to the inflationary Hubble scale
$H$ still presents a severe problem for generic supergravity
theories. Remarkably though, it turns out that among the class of
supergravity models emerging from
orbifold compactifications
of superstring theory, one can find a large subset for which
this problem disappears.
As a toy model, we will see how things work out with a specific
choice for the perturbative part of the superpotential.

Another crucial difference concerns the mass $m$. In contrast with the
true vacuum case, the cmb anisotropy does not determine $m$
in the vacuum dominated case,
but rather
determines $V_0$ as a function of $m$.
The value $m\sim
10^{13}\,$GeV that obtains in the true vacuum case is allowed as an
upper limit, but $m$ can be almost arbitrarily small and
it is natural to contemplate values
down to at least the scale $m\sim 100\,$GeV.
The value of $m$ chosen by nature might be accessible
to observation because it determines the spectral index $n$; if
$m$ is within an order of magnitude or so of its upper limit
$n$ is appreciably higher than 1, whereas
if it is much lower $n$ is indistinguishable from 1.
In the superstring motivated models mentioned earlier,
the first case probably obtains if the
slope of the inflationary potential is dominated by
one-loop corrections coming from the
Green-Schwarz mechanism, in which case the value of $n$ is determined
by the orbifold. This would open up the interesting possibility
that observations of the cmb anisotropy and large scale structure
provide a window on superstring physics.

The opposite case $m\sim100\,$GeV is also interesting.
Supersymmetric theories of particle physics typically contain several
scalar fields with this mass. The corresponding false vacuum energy
scale $V_0^{1/4}\sim 10^{11}\,$GeV also appears in particle physics,
as that associated with Peccei-Quinn symmetry, a global $U(1)$
symmetry which is perhaps the most promising explanation for
the observed CP invariance of the strong interaction. This same symmetry
provides the axion, which is one of the leading dark matter candidates,
and the possibility that it might in addition
provide the false vacuum for inflation is to say the least interesting.
We explore this possibility in the context of global supersymmetry and
find that it can easily be realised there. We have not gone on to explore it
in the context of supergravity, but there seems to be no reason why
it should not be realised within the context of the superstring derived
models considered earlier.

As will be clear from this introduction, the present work
is expected to be of interest to a very wide audience, ranging
from observational astronomers to superstring theorists. With this in
mind we have tried to keep separate the
part of the paper that discusses the phenomenology of
the false vacuum inflation models, and the part that
relates these models to particle physics.

The outline of the paper is as follows. Section \ref{INFL} introduces the
specific second-order model upon which most of our discussion shall be
focussed. We analyse the inflationary dynamics and density perturbation
constraints by a combination of analytic and numerical methods to delineate
the observationally viable models. Section \ref{TOP} then takes our attention
onto the formation of topological defects, which (almost) inevitably form at
the end of inflation. Their possible existence constrains the models, and
there is the further opportunity of a reconciliation of structure-forming
defects with inflation. In Section \ref{SUGR} we try to realise the model
in the contexts of global supersymmetry, supergravity and superstring derived
supergravity. In Section \ref{MODELS} we consider the related first-order
model which also indicates the link with extended inflation models.
Section \ref{CONC} summarises the paper.

\section{Inflationary Phenomenology}
\label{INFL}
\setcounter{equation}{0}
\def\theequation{\thesection.\arabic{equation}}

\subsection{The Model}

Throughout this paper we assume Einstein gravity. During inflation the energy
density is supposed to be dominated by the potential of two scalar fields,
which is taken to be of the form
\begin{equation}
\label{FULLPOT}
V(\phi,\psi) = \frac{1}{4} \lambda \left( \psi^2 - M^2 \right)^2
	+ \frac{1}{2} m^2 \phi^2 + \frac{1}{2} \lambda' \phi^2 \psi^2 \,.
\end{equation}
This potential possesses the symmetries $\phi \leftrightarrow -\phi$ and $\psi
\leftrightarrow -\psi$, and is the most general renormalisable potential with
this property except for a quartic term $\lambda'' \phi^4$.\footnote{The pure
quadratic term is the simplest possibility, and is also the one favoured by
particle physics considerations (Section \ref{SUGR}). Non-renormalisable
potentials, involving higher powers of the fields, arise naturally in the
context of supergravity, but for simplicity we ignore them here. As we discuss
later the $\psi$ field can have several components, but they do not affect the
issues we discuss in the present section.}

We make the restrictions $0< \lambda$, $\lambda' \lsim 1$, and we also require
that the masses $m$ and $M$ fall in the range between $100\,$GeV and the
Planck scale $\mpl/\rep=2.4 \times 10^{18}\,$GeV indicated by particle physics
considerations.\footnote{The factor $\rep$ is mathematically convenient and we
shall follow the majority of authors by inserting it, though of course our
understanding of the Planck scale is quite insufficient to justify such
factors from a physical viewpoint.}

Provided that $\phi^2 > \phi_{{\rm inst}}^2 $, where
\begin{equation}
\phi_{{\rm inst}}^2 = \lambda M^2/\lambda' \,,
\end{equation}
there is a local minimum at $\psi=0$ on the constant $\phi$ slices,
corresponding to a false vacuum. Our assumption is that  inflation occurs with
the $\psi$ field sitting in this false vacuum, so that the potential is
\begin{equation}
\label{PHIPOT}
V(\phi) = \frac{1}{4} \lambda M^4 + \frac{1}{2} m^2 \phi^2 \,.
\end{equation}
If the false vacuum dominates, inflation ends when $\phi$ falls below
$\phi_{{\rm inst}}$, the fields rapidly adjusting to their true vacuum values
$\psi=M$ and $\phi=0$.

This model was first considered by Kofman and Linde \cite{Kofman}, who pointed
out that it might produce cosmic strings with enough energy per unit length to
form structure. They considered only what we shall term
the inflaton dominated regime (small false vacuum energy), as did several
subsequent authors studying this and related
models \cite{vishniac,kofpog,yokoyama,kbhp,sbb,lyth90,hodpri,nagasawa}.
In
order to obtain interesting effects, these authors had to assume (at least)
that
the coupling $\lambda'$ was many orders of magnitude less than unity. The case
of false vacuum domination, which is our main focus, was proposed by Linde who
termed it `Hybrid Inflation' \cite{LIN2SC} and has
received further attention from
Liddle and Lyth \cite{LL2}, Linde \cite{LIN2SC2}, and Mollerach, Matarrese and
Lucchin \cite{MML}. In this case the couplings can be of order unity,
but for completeness we explore also the regime of parameter space where
they are very different from one.

\subsection{Inflationary dynamics}
\label{infldyn}

As usual, the inflationary dynamics are governed by the equations
\begin{eqnarray}
H^2 & = & \frac{8\pi}{3 \mpl^2} \left( \frac{1}{2} \dot{\phi}^2 +
	\frac{1}{2} \dot{\psi}^2 + V(\phi,\psi) \right) \,,\\
\ddot{\phi} + 3 H \dot{\phi} & = & - \frac{\partial V(\phi, \psi)}{\partial
	\phi} \,,\\
\ddot{\psi} + 3 H \dot{\psi} & = & - \frac{\partial V(\phi, \psi)}{\partial
	\psi} \,,
\end{eqnarray}
for two isotropic scalar fields in an expanding universe, with $H = \dot{a}/a$
the Hubble parameter, $a$ the scale factor, $\mpl$ the Planck mass and dots
derivatives with respect to time. Our assumption is that there is a transitory
regime during which the $\psi$ field rolls to $\psi=0$ from whatever its
initial value may have been, and is followed by sufficient inflation on
the $\psi=0$ trajectory to erase any evidence of such a transient.
Inflation then proceeds according to the usual single field equation for
$\phi$ in the potential of Eq.~(\ref{PHIPOT}).
Without loss of generality, we shall assume that $\phi$ is initially positive.

We shall utilise the slow-roll approximation throughout. It is characterised
by the conditions
\begin{equation}
\epsilon \ll 1 \quad ; \quad |\eta| \ll 1 \label{epssmall} \,,
\end{equation}
where the two dimensionless functions
$\epsilon(\phi)$ and $\eta(\phi)$ are defined by
\begin{eqnarray}
\epsilon(\phi) & \equiv &\frac{\mpl^2}{16\pi} \left(
	\frac{V'(\phi)}{V(\phi)} \right)^2 \,,\\
\eta(\phi) & \equiv & \frac{\mpl^2}{8\pi} \frac{V''(\phi)}{V(\phi)} \,.
\end{eqnarray}
Here and throughout primes indicate derivatives with respect to the field
$\phi$. With
justification from numerical results, it is standard to assume that if
the potential satisfies
these conditions, then the
solutions for a broad range of initial conditions rapidly approach the
attractor
\begin{equation}
\label{ATTR}
3 H \dot{\phi} \simeq - V' \,.
\end{equation}
When this is satisfied, there exists a simple expression for the number
$N$ of $e$-foldings of expansion which occur between two scalar field
values $\phi_1$ and $\phi_2$
\begin{equation}
N(\phi_1,\phi_2) \equiv \ln \frac{a_2}{a_1} \simeq - \frac{8 \pi}{\mpl^2}
	\int_{\phi_1}^{\phi_2} \frac{V}{V'} \, {\rm d}\phi \,.
\end{equation}
For our specific potential we have
\begin{eqnarray}
\label{ETA}
\eta & = & \frac{m^2 \mpl^2}{2\pi \left( \lambda M^4 + 2 m^2 \phi^2
	\right)} \,,\\
\label{EPS}
\epsilon & = & \frac{\mpl^2 m^4 \phi^2}{\pi \left( \lambda M^4 + 2 m^2
	\phi^2 \right)^2}  =  \frac{1}{2} \mpsinv \eta^2\phi^2 \,,\\
\label{EFOLD}
N(\phi_1,\phi_2) & = & \frac{2\pi \lambda M^4}{m^2 \mpl^2} \ln
	\frac{\phi_1}{\phi_2} + \frac{2\pi}{\mpl^2} \left( \phi_1^2 -
	\phi_2^2 \right) \,.
\end{eqnarray}

Within the slow-roll approximation, the condition for inflation to occur is
simply that $\epsilon$ be less than one. However, slow-roll is automatically a
 poor approximation should $\epsilon$ reach this value, though the amount of
inflation that occurs as $\epsilon$ becomes large is always small. Numerical
simulation indicates that for this potential if $\epsilon$ and $\eta$ grow to
unity, shortly thereafter the inflationary condition  $\ddot{a} > 0$ is
violated and inflation ends. The number of  $e$-foldings that occur between
these events is a tiny fraction of unity, and can be ignored. It is therefore
sensible operationally to identify the end of inflation  in this case with the
precise condition that $\epsilon = 1$, should this  occur, and we shall assume
this subsequently.

There are therefore two separate ways in which inflation may end in
this model, the one which is applicable depending on the parameter
values. These are
\begin{enumerate}
\item If $\phi$ reaches $\phi_{{\rm inst}}$ while inflation is occurring, then
inflation may end through the instability of the $\psi$ field to roll to its
global minimum. As noted by Linde \cite{LIN2SC2} one expects this to happen,
at least for $\lambda$ and $\lambda'$ not too small, if the false vacuum term
$\lambda M^4/4$ dominates the potential. We look in some detail at this
possibility in Section \ref{TOP}, confirming the picture of rapid
instability.
\item If the logarithmic slope of the potential becomes too large {\em on the
$\psi = 0$ trajectory}, then inflation can end while the $\phi$ field is still
rolling down that trajectory. This is symptomised by $\epsilon$ growing to
exceed unity. Some time later, $\phi$ will pass $\phi_{{\rm inst}}$ and the
$\psi$ field may roll away from $\psi = 0$.
\end{enumerate}

The value of $\phi$ at which $\epsilon$ becomes equal to unity
is\footnote{There is also a second root at smaller $\phi$, where $\epsilon$
drops back below unity. However, for second-order models it is easy to show
that the attractor solution Eq.~(\ref{ATTR}) cannot be attained for $\phi$
below this root, allowing inflation to restart, before the instability sets
in.}
\begin{equation}
\phi_\epsilon = \frac{\mpl}{\sqrt{16\pi}} \left( 1 + \sqrt{1 - \mpsinv
	\frac{\lambda M^4}{m^2}} \right) \,.
\label{phiep}
\end{equation}
If $8\pi \lambda M^4/\mpl^2 m^2 > 1$, then $\phi_\epsilon$ does not exist at
all. In that case inflation must end by instability. In the opposite limit,
the position $\phi_\epsilon\to \mpl/\sqrt{4\pi}$ is familiar from chaotic
inflation with a single field, and of course the standard results will be
recovered in that limit with the $\psi$ field playing no significant role.

We need to know the number $N(k)$ of
Hubble times of inflation which occur after
a given scale leaves the horizon\footnote
{As usual we say that a comoving scale
$a/k$ leaves the horizon when $aH/k=1$}.
With the assumptions
(valid in our model) that $H$ does not vary significantly and that reheating
is prompt it is given by \cite{LL2}
\begin{equation}
N(k)= 62- \ln \frac{10^{16}\,\mbox{GeV}}{V_1^{1/4}}
-\ln \frac {k}{a_0 H_0} \,,
\end{equation}
where subscript `0' indicates present value.
The largest cosmologically interesting scale is of order the
present Hubble distance
(roughly the size of the observable universe), $k = a_0 H_0$, and other
scales of cosmological interest leave
the horizon at most a few Hubble times after this one.
As for the inflationary energy scale,
true vacuum inflation typically gives $V_1^{1/4}\sim 10^{16}$\,GeV,
which makes the observable universe leave the horizon about 60 e-folds
before the end of inflation (the fact
that reheating may be very inefficient in this
model may reduce this number somewhat). As we shall see,
false vacuum dominated inflation can give
values as low as $V_1^{1/4}\sim 10^{11}\,$GeV, which reduces the figure
60 to about 50.
However, one only needs a rough estimate of $N$ for most
purposes because the potential is slowly varying, and for simplicity
we suppose from now on that cosmologically interesting scales leave
the horizon 60 e-folds before the end of inflation.

Provided the
parameters are chosen in such a way as to produce the correct level
of density perturbations to explain the COsmic Background Explorer (COBE)
satellite observations \cite{COBE} of the
cmb anisotropies, there
is no problem in obtaining sufficient inflation to resolve the horizon and
flatness problems or with ensuring that a classical description of the
evolution is adequate. We thus need only investigate the density perturbation
constraint in order to completely fix the model.

\subsection{Density perturbations}
\label{dperts}

The adiabatic density perturbation, which is generally thought to be
responsible for large scale structure, originates as a vacuum fluctuation
during inflation.
Its spectrum is determined by a
quantity $\delta_H$, which loosely speaking gives the density contrast at
horizon crossing and is defined formally in
\cite{LL2}. The inflationary prediction for the spectrum is
\begin{equation}
\delta_H^2 (k) = \frac{32}{75} \frac{V_*}{\mpl^4} \frac{1}{\epsilon_*} \,,
\label{delhone}
\end{equation}
where $\epsilon$ is the slow-roll parameter defined earlier and the subscript
$*$ indicates that the right hand side is to be evaluated as the comoving
scale $k$ equals the Hubble radius ($k = aH$) during inflation.

By virtue of the slow-roll conditions
Eqs.~(\ref{epssmall})-(\ref{ATTR}), this formula gives a value of
$\delta_H(k)$ which is nearly independent of $k$ on scales of
cosmological interest, in agreement with observation.
For a sufficiently flat spectrum, and provided that no significant generation
of long wavelength gravitational wave modes occurs, the central value of the
COBE $10^\circ$ anisotropy, $30 \mu K$, is reproduced provided
one has
$\delta_H = 1.7 \times 10^{-5}$ \cite{LL2}\footnote{This figure assumes a
Gaussian beam profile, and is raised by 16\% if the precise profile of
the experiment is used and a correction applied for the incomplete
sky coverage inducing errors in the monopole and dipole subtractions
\cite{newcobe}. For an accurate
analysis one has to include (here and elsewhere) the effect of
spectral tilt and gravitational waves on the COBE normalisation
\cite{LiddleES}. Such changes are
not significant in the present context except for extreme parameter
values, and for simplicity we shall not include
them in the normalisation though we shall discuss tilt and gravitational
waves later.}.
Thus the inflationary energy scale when cosmologically interesting
scales leave the horizon is given by
\begin{equation}
V_{60}^{1/4}=6 \epsilon_{60}^{1/4} \times 10^{16}\,\mbox{GeV} \,,
\label{infenergy}
\end{equation}
where a subscript 60 denotes 60 e-folds before the end of inflation.

The most efficient way to proceed is as follows. First, fix the couplings
$\lambda$ and $\lambda'$. Then, having chosen a value for the mass scale $M$,
find the value(s) of $m$ such that the density perturbation
constraint is satisfied. Assuming that inflation ends promptly if $\phi$ falls
below $\phi_{\rm inst}$, we can determine the means by which inflation ends and
the corresponding value of $\phi$
\begin{equation}
\phi_{{\rm end}} = \max \{\phi_\epsilon,\phi_{{\rm inst}} \} \,.
\end{equation}
We then use Eq.~(\ref{EFOLD}) to determine the value of $\phi$ 60
$e$-foldings from the end of inflation, $\phi_{60}$, and
evaluate $\delta_H$ as
\begin{equation}
\label{DELH}
\delta_H^2 = \frac{8\pi}{75} \frac{\left( \lambda M^4+2m^2\phi_{60}^2
	\right)^3}{\mpl^6 m^4 \phi_{60}^2} \,.
\end{equation}
To find the value(s) of $m$ which satisfy the COBE normalisation,
remember
that $\phi_{{\rm end}}$, and hence $\phi_{60}$, is a function of $m$. In
general this procedure cannot be carried out analytically, and we compute
using an iterative numerical method. However, the problem can be solved
analytically and self-consistently in two regimes. As we shall see, provided
$M$ is not too large then for each $M$ there are {\em two} possible choices
of $m$ which give the right perturbation amplitude. One corresponds to the
traditional polynomial chaotic inflation scenario, where the first term in
Eq.~(\ref{PHIPOT}) plays a negligible role (and by implication the first term
in the numerator of Eq.~(\ref{DELH}) likewise). [There is a variant on this
regime, also discussed below, where the instability sets in while the false
vacuum energy is still
negligible.]
The second, and for our purposes more interesting, possibility
involves a value of $m \ll M$, and corresponds to domination by the first term
in Eq.~(\ref{PHIPOT}).

\subsection{Delineating parameter space}

We shall now examine different analytic and numerical regimes. The results are
concisely summarised in Figure 1.

\subsubsection*{The inflaton dominated regime}

The simplest scenario of all is one in which the $\psi$ field plays no role
whatsoever, leaving just the $\phi$ field to govern inflation. The potential
$V=m^2\phi^2/2$ was proposed by Linde \cite{CHAOTIC} as a simple realisation
of chaotic inflation. With this potential, inflation ends when $\phi$ starts
to oscillate around its minimum, which occurs when $\epsilon \simeq 1$
corresponding to
\begin{equation}
\phi_{{\rm end}} \simeq \mpl/\sqrt{4\pi} \,.
\label{phiend}
\end{equation}
The condition that our potential Eq.~(\ref{PHIPOT}) be a good approximation to
this one is therefore
\begin{equation}
\mpsinv \frac{\lambda M^4}{4 m^2} \ll 1 \,.
\label{chaotic}
\end{equation}
In Eq.~(\ref{EFOLD}) the second term dominates, giving
\begin{equation}
\phi_{60}
\simeq \sqrt{\frac{60}{2\pi}} \, \mpl \,.
\end{equation}
Note that in this regime the characteristic scale of $\phi_{60}$ is the Planck
scale. The density perturbation amplitude is independent of $M$,
$\lambda$ and $\lambda'$ in this limit, and the correct value is obtained with
\begin{equation}
\frac{\rep}{\mpl} m = \frac{\pi}{4\sqrt6} \delta_H =5.5 \times 10^{-6} \,.
\label{mchaotic}
\end{equation}
The condition for the validity of this approximation is therefore
\begin{equation}
\frac{\sqrt{8\pi}}{\mpl}
\lambda^{1/4} M \ll 3\times 10^{-3}
\,.
\label{Mchaotic}
\end{equation}

The above analysis assumes that $\phi>\phi_{\rm inst}$
throughout inflation, which from
Eq.~(\ref{phiend}) fails if $\phi_{{\rm inst}}\gsim \mpl/\sqrt{4\pi}$, or
equivalently
\begin{equation}
\frac{\lambda'^2}{\lambda} \lsim \frac14 \lambda M^4 \left( \mpsinv
	\right)^2 \lsim 10^{-11} \,.
\end{equation}
(The final inequality is Eq.~(\ref{Mchaotic}).) If $\phi$ falls below
$\phi_{{\rm inst}}$, then as discussed in Section~3.1 $\psi$ may roll towards
its minimum at fixed $\phi$,
\begin{equation}
\psi^2_{{\rm vac}}(\phi) = M^2 \left( 1 - \frac{\phi^2}{\phi_{{\rm inst}}^2}
	\right) \,.
\label{psivac}
\end{equation}
It oscillates around the minimum, losing energy through the expansion of the
universe so that after a few Hubble times $\psi\simeq\psi_{\rm vac}$ (if its
spatial gradient is not negligible it may settle down more quickly through
thermalisation). Inserting $\psi=\psi_{\rm vac}$ into Eq.~(\ref{FULLPOT})
gives \cite{Linde92}
\begin{eqnarray}
V(\phi) & = & \frac{1}{2} m^2 \phi^2 + \frac{\lambda}{4} M^4 \left[1-
	\left(1-\frac{\phi^2}{\phi_{{\rm inst}}^2}\right)^2\right] \,,
	\label{vnew} \\
V'(\phi) & = & m^2 \phi
	\left[ 1 + \frac{\lambda M^4}{m^2\phi_{{\rm inst}}^2}
	\left( 1 - \frac{\phi^2}{\phi_{{\rm inst}}^2}\right)\right] \,,
\label{vpnew}\\
V''(\phi) & = & m^2 \left[1+\frac{\lambda M^4}{m^2\phi_{{\rm inst}}^2}
	\left( 1 - 3 \frac{\phi^2}{\phi_{{\rm inst}}^2}\right)\right] \,.
\label{vppnew}
\end{eqnarray}
Since we are in the regime $\phi_{{\rm inst}}\gsim \mpl/ \rep$, the condition
Eq.~(\ref{chaotic}) written down earlier guarantees again that the
modification to $V$ will be negligible.

It therefore appears that when Eq.~(\ref{chaotic}) is well satisfied, the
evolution of $\phi$ will not be significantly affected even if $\phi$ falls
below $\phi_{{\rm inst}}$. If Eq.~(\ref{chaotic}) is only marginally
satisfied, the evolution of $\phi$ might be substantially altered, leading to
a significant change in the predicted adiabatic density perturbation. The
simplest assumption is that the potential is given by Eq.~(\ref{vnew}). In
that case, if one ignores the inhomogeneity of $\phi$ caused by the phase
transition, the perturbation is still given by the usual formula,
Eq.~(\ref{delhone}), with the new potential \cite{Linde92}. However, this
formula depends crucially on the assumption that each Fourier mode of $\phi$
is in the vacuum state before leaving the horizon, whereas the phase
transition will inevitably populate some of the modes with non-zero particle
number. Taking this into account, the adiabatic perturbation on scales leaving
the horizon after $\phi=\phi_{{\rm inst}}$ might be quite different
(non-Gaussian, with a non-flat spectrum and a different normalisation).
An additional adiabatic perturbation might also be generated
by the perturbation in $\psi$ \cite{kofpog,kbhp,sbb}, as discussed in
Section 3.1.

\subsubsection*{The vacuum dominated regime}

We now explore the opposite regime, where the vacuum energy density notionally
associated with the $\psi$ field dominates the potential. Special cases of
this regime of parameter space have already been considered in
\cite{LIN2SC}--\cite{MML}.
%%,LIN2SC2,LL2,MML}.

As noted earlier, inflation is expected to occur only if $\eta$ and $\epsilon$
are small compared with unity. The first of these parameters is independent of
$\phi$ in the limit of vacuum domination, with the value
\begin{equation}
\eta = \mps \frac{4m^2}{\lambda M^4} \,.
\end{equation}
Thus, the requirement that $\eta\lsim 1$ in this regime is precisely the
opposite of the condition Eq.~(\ref{chaotic}) which characterises the regime
in which the vacuum does {\it not} dominate. The parameter $\epsilon$ {\em
decreases} as inflation proceeds, and during the era $\phi<\phi_{60}$ that we
are interested in we have
\begin{equation}
\epsilon < \epsilon_{60} = \frac12 \mpsinv \eta^2 \phi_{60}^2 \,.
\label{epssixty}
\end{equation}
The condition for vacuum domination is
\begin{equation}
\frac12 \mpsinv \phi_{60}^2 \eta \ll 1 \,,
\label{vacdom}
\end{equation}
or equivalently
\begin{equation}
\epsilon_{60} \ll \eta \,.
\label{vdomfirst}
\end{equation}

The first term of Eq.~(\ref{EFOLD}) dominates the formula for $\phi_{60}$,
giving
\begin{equation}
\phi_{60}^2 = \frac{\lambda}{\lambda'} M^2 e^{120\eta} \,.
\label{phisixty}
\end{equation}
The COBE normalisation Eq.~(\ref{DELH}) is therefore
\begin{eqnarray}
\label{XXXX}
\frac{\rep}{\mpl} \sqrt{\lambda'}M & = & 10\sqrt3 \pi
	\delta_H \eta e^{60\eta} \,,\\
 & = & 9.3 \times10^{-4} \eta e^{60\eta} \,.
\label{cmb}
\end{eqnarray}
It involves the two masses and the two coupling constants only in the
dimensionless combinations
\begin{eqnarray}
\label{Mhat}
\hat{M} & \equiv &  \frac{\rep}{\mpl} \sqrt{\lambda'} M \,,\\
\label{mhat}
\hat{m} & \equiv & \frac{\rep}{\mpl} \sqrt\frac{\lambda'^2}{\lambda} \, m \,,
\end{eqnarray}
since $\eta=4\hat{m}^2/\hat{M}^4$. The restrictions $M\lsim \mpl/\rep$ and
$\lambda'\lsim 1$ that we have agreed to impose because of particle physics
considerations means that we are in the regime $\hat{M}\lsim 1$, which
with the COBE constraint corresponds to $\eta \lsim 0.15$.

The situation becomes especially simple in the regime $60\eta\ll 1$, which we
call the {\it extreme vacuum dominated regime}. It
corresponds to $\hat M \ll 10^{-4}$, and
the quantity $\eta$ varies linearly with
$\hat M$ leading to \cite{LIN2SC,LIN2SC2}
\begin{equation}
\frac {\hat M^5}{\hat m^2}
	= 40\sqrt3 \pi \delta_H = 3.7 \times 10^{-3} \,.
\label{cmbconstraint}
\end{equation}
Inserting the Planck mass and working with the masses themselves
this formula becomes
\begin{equation}
\frac{M}{5.5\times 10^{11}\,\mbox{\rm GeV}}
=\lambda'^{-1/10}\lambda^{-1/5}
\left( \frac{m}{1\,\mbox{\rm TeV}} \right)^{\frac25} \,.
\label{extvacdom}
\end{equation}
In the other regime $60\eta \gsim 1$ \cite{LL2,MML}, $\eta$ varies only
logarithmically with $M$, and the power $m^{2/5}$ gradually changes to
$m^{1/2}$.

Although the cmb constraint can be expressed in terms of just the two
quantities $\hat{m}$ and $\hat{M}$,
the vacuum domination condition involves three quantities which are
conveniently chosen to be $m$, $\lambda^{1/4} M$ and $\lambda'^2/\lambda$. It
is therefore useful to express the cmb constraint as a constraint on $m$ and
$\lambda^{1/4}M$ at fixed $\lambda'^2/\lambda$. (Another good reason for doing
this is that $\lambda M^4$ is the false vacuum energy density.)
The extreme vacuum dominated regime $60\eta\ \ll 1$ corresponds to
\begin{equation}
\lambda^{1/4} M \ll 4 \times 10^{-5}
\left(\frac{\lambda'^2}{\lambda}\right)^{-1/4} \frac{\mpl}{\rep}
\,.
\end{equation}
In this regime $\eta$
increases linearly with $\lambda^{1/4} M$,
\begin{equation}
\frac\eta4 \equiv \mps \frac{m^2}{\lambda M^4} \simeq 270 \left(
	\frac{\lambda'^2}{\lambda}\right)^{1/4} \frac{\rep}{\mpl}
	\lambda^{1/4} M \,,
\label{small}
\end{equation}
while for larger values it increases only logarithmically giving the
normalisation
\begin{equation}
\label{large}
\frac{\mpl^2}{8\pi} \, \frac{m^2}{\lambda M^4} \simeq 0.004
\mbox{\ to\ } 0.04
\end{equation}
with the upper limit corresponding to $\eta = 0.15$.

We have yet to invoke the false vacuum domination condition
Eq. (\ref{vacdom}). Using Eq.~(\ref{cmb}), it becomes
\begin{equation}
\eta^3 e^{240\eta}\ll 2\times 10^6 \frac{\lambda'^2}{\lambda} \,.
\end{equation}
With ${\lambda'}^2/\lambda = 1$, this bound is saturated for
$\eta=0.09$, and a similar limit is obtained
for any value of the ratio within a few orders of magnitude of
unity. Setting $\eta=0.09$, one learns that the false vacuum dominated
regime is restricted to
\begin{eqnarray}
\label{Mhighlim}
\frac{\rep}{\mpl} \lambda^{1/4} M & \lsim & 2\times 10^{-2}
	\left(\frac{\lambda'^2}{\lambda}\right)^{-1/4} \,, \\
\frac{\rep}{\mpl} m & \lsim & 8 \times 10^{-5}
	\left(\frac{\lambda'^2}{\lambda}\right)^{-1/2}
\label{mhighlim} \,.
\end{eqnarray}

The upper limit on $\lambda^{1/4} M$ is not far below the
one following just from the
fact that $\epsilon_*<1$ in the cmb constraint Eq.~(\ref{delhone}), which
using $\frac14\lambda M^4<V$ is\footnote
{To understand the relation between the two limits, note that
when the vacuum domination condition Eq.~(\ref{vacdom}) is saturated,
$V=\frac12\lambda M^4$ and $\epsilon_{60}=\eta$.}
\begin{equation}
\frac{\rep}{\mpl} \lambda^{1/4} M < 5 \times 10^{-2} \,.
\label{Mabs}
\end{equation}

\subsubsection*{The intermediate regime}

We have now investigated the extreme cases, first the one in which the
false vacuum energy is negligible, and second the one in which it
dominates. There remains the intermediate case where both are
comparable, during at least part of the cosmologically significant
era $\phi<\phi_{60}$.

Plotted with $m$ the vertical axis and $\lambda^{1/4}M$ the horizontal axis,
we have learned that the first regime corresponds to a straight horizontal
line, whereas the second one corresponds to a line with positive slope. Unless
$\lambda'^2/\lambda$ is several orders of magnitude away from unity,
comparison of Eqs.~(\ref{mchaotic}) and (\ref{Mchaotic}) with
Eqs.~(\ref{mhighlim}) and (\ref{Mhighlim}) shows that the right hand ends of
these two lines are separated by at most an order of magnitude or so in the
$m$ and $\lambda^{1/4} M$ variables. Therefore the intermediate regime is
not very extensive, but it is still important to investigate it in order to
see if new physics occurs.

Even in the intermediate regime the upper bound Eq.~(\ref{Mabs}) holds. Apart
from this fact, numerical techniques are required to solve the density
perturbation constraint. The solution, as might be expected, is that as
$\lambda^{1/4}M$ is increased the two solution branches approach each other
and merge continuously. This merger specifies the maximum allowed value of
$\lambda^{1/4}M$; for higher values it becomes impossible to obtain a
sufficiently low perturbation amplitude regardless of the choice of $m$. (The
maximum value does depends on $\lambda$ and $\lambda'$, of course.) Figure 1
illustrates the complete set of viable models for the couplings both set to
unity, showing both the asymptotic regimes and the merger region.

\subsection{Tilt and gravitational waves}
\label{tilt}

Although the inflationary prediction in Eq.~(\ref{delhone}) for the spectrum
$\delta_H(k)$ is almost flat, there is always some
$k$-dependence, usually referred to in the literature as
{\it tilt}. On cosmologically interesting scales the tilt
can usually be well characterised by a constant spectral index $n$,
such that $\delta_H^2\propto k^{n-1}$, and in that case one learns from
the slow-roll conditions Eqs.~(\ref{epssmall})-(\ref{ATTR})
that \cite{LL1,LL2}
\begin{equation}
n-1=2\eta_{60}-6\epsilon_{60} \,.
\label{index}
\end{equation}
As always, we take `cosmologically interesting' to mean scales that leave the
horizon 60 e-folds before the end of inflation.

In addition to the adiabatic density (scalar) perturbation, inflation
also generates gravitational waves, whose contribution
$R$ to the cmb anisotropy $(\Delta T/T)^2$
relative to that of the scalar modes is \cite{starob,LL1,LL2}
\begin{equation}
R\simeq 12\epsilon_{60} \,.
\end{equation}

For true vacuum inflation with a  potential $V\propto e^{A\phi}$,
$\eta=2\epsilon$ so that $n$ is less than 1 and $R\simeq 6(1-n)$.
Replacing the exponential by a power $\phi^\alpha$ gives tilt
$n-1=-(2+\alpha)/120$, still negative, and
provided that $\alpha\geq2$ as required by particle physics
it still gives $R\simeq 6(1-n)$. Thus, true vacuum inflation with
a $\phi^\alpha$ potential typically makes
$n$ a few percent below unity and it makes $R$ tens of percent.
Both of these predictions are big
enough to be cosmologically significant.

The case of false vacuum dominated
inflation is dramatically different. The condition Eq.~(\ref{vacdom})
for false vacuum
domination is $\epsilon\ll\eta$, and unless it is almost saturated
$\eta$ is very small. As a result, the
tilt and gravitational wave contribution are
both indistinguishable from zero, for generic choices of the parameters.
For fixed couplings, they
become significant only at the upper end of the mass range allowed by
Eq.~(\ref{vacdom}), and in contrast with the true vacuum case
$n$ is {\it greater} than one until Eq.~(\ref{vacdom}) is almost saturated.

The value of $n$ for a given parameter choice
is obtained by substituting
Eq.~(\ref{phisixty}) into Eq.~(\ref{index}).
With $\lambda'^2/\lambda=1$,
$n$ is equal to $1.0001$ for the minimum value $m\sim 100$\,GeV
at the lower end, and rises to a maximum value
$n=1.14$ near the upper end of the allowed range
\cite{LL2}.
The biggest possible value of $n$, corresponding to $\eta=0.15$
and $\epsilon\ll 1$, is $n=1.30$, but this is only achieved with
extreme values of the parameters $M\sim \mpl/\rep$ and
${\lambda'}^2/\lambda\gsim 10^{9}$.

We have extended these results numerically to the intermediate regime
for the case where the couplings are unity.
The result for $n$ is shown in Figure 2
as a function of $\lambda^{1/4} M$. The two solution branches are as in Figure
1. The analytic vacuum dominated result
agrees well with the exact one until well beyond the maximum, and in
particular the maximum value $n \simeq 1.14$ is essentially the same.
This number is considerably less than that which was suggested
could be obtained in this model by other authors \cite{LIN2SC2,MML};
they
extrapolated the vacuum dominated case beyond its regime of validity
and neglected $\epsilon$ to obtain
their larger values. As we commented above, this conclusion is not altered
unless the couplings are changed (and separated) by orders of magnitude.

Another interesting feature is the dip in $n$ to around 0.92 as one exits from
the inflaton dominated regime into the intermediate region, indicating that
these models are also capable of providing a significant (though not
startling) tilt in the opposite direction. In the limit of small $M$,
the vacuum dominated case asymptotes to unity and the inflaton dominated case
to the standard value 0.97.

We have also calculated the gravitational wave component, though we have not
attempted to include it (or the tilt) into the COBE normalisation.
Gravitational waves make only a small contribution to COBE except in the
intermediate regime where they can reach a peak of tens of percent (though as
expected when $n$ exceeds unity the gravitational wave component is suppressed
by the small $\epsilon$ required), indicating
that a proper treatment of them is required to develop the
precise phenomenology of the intermediate regime.

\section{The Second-Order Phase Transition}
\label{TOP}
\setcounter{equation}{0}
\def\theequation{\thesection.\arabic{equation}}

If $\phi$ falls below $\phi_{{\rm inst}}$ before the end of inflation, the
false vacuum is destabilized and there is a possibility of a second-order
phase transition, of a kind quite different from the usual thermal phase
transition. In this section we consider the nature of this transition,
treating separately the very different regimes of inflaton domination and
vacuum domination.

For simplicity we continue to suppose that $\psi$ is a single real field with
the potential Eq.~(\ref{FULLPOT}). This potential has the discrete symmetry
$\psi \leftrightarrow -\psi$, which of course implies that the phase
transition creates domain walls located at surfaces in space where $\psi$
vanishes. One can instead have $N$ real fields, and replace $\psi^2$ by $\sum
|\psi_i|^2$ in the potential Eq.~(\ref{PHIPOT}), which has $O(N)$ symmetry.
This will give global strings ($N=2$), global monoples ($N=3$) or textures
($N\geq 4$) if the symmetry is global, or gauge strings ($N=2$) or gauge
monopoles ($N=3$) if it is local. We expect that in all cases our
discussion of the
evolution of $\phi$ and $\psi$ should be roughly correct, provided that $\psi$
is taken to represent the `radial' degree of freedom with respect to which the
potential has a maximum as opposed to the `angular' degrees of freedom with
respect to which it is constant.

One other significance of having a local symmetry rather than a global one,
emphasised by Linde \cite{LIN2SC2}, is that one might have no defects forming
simply because the lowest homotopy groups all vanish. This takes advantage of
there being no such thing as local texture or nontopological texture, due to
the gauge degrees of freedom cancelling the scalar gradients. For global
symmetries however, scalar gradients can still play a
harmful role even in the absence of topological constraints.

We will see that the formation of topological defects at the end of a period
of inflation offers the intriguing possibility that we could make use of the
inflationary epoch to solve the flatness issues of our universe, and yet
retain the possibility of utilising defects as the source of the density
fluctuations to seed large scale structure.

\subsection{The inflaton dominated regime}
\label{secphidom}

In Section 2.4, we saw that in the inflaton dominated regime the false vacuum
is
maintained right up to the end of inflation unless $\lambda'$ is very small.
In that case the phase transition to the true vacuum will take place after
inflation, and will presumably be of the usual thermal type, any topological
defects forming by the usual Kibble mechanism \cite{twb76}.

If $\lambda'$ is sufficiently small, $\phi$ can fall below $\phi_{\rm inst}$
before the end of inflation. Depending on the regime of parameter space, the
transition to the false vacuum and the formation of defects may then occur
before the end of inflation. This case has been treated by several
authors\footnote{Some of these authors consider a coupling of $\psi$ to the
spacetime curvature $R$ rather than to the inflaton field, but this is
equivalent for the present purpose.}
\cite{Kofman,vishniac,kofpog,yokoyama,kbhp,sbb,lyth90,hodpri,nagasawa},
and we
look briefly at the results of these authors because they provide a starting
point for our discussion of the vacuum dominated case, which is our main
focus. As we noted in Section~2.4, the back reaction of $\psi$ on $\phi$ is
negligible in the vacuum dominated regime. As a result we can in
principle follow the evolution of $\psi$ explicitly, using quantum field
theory in curved spacetime
\cite{vishniac,kofpog,yokoyama,kbhp,sbb,lyth90,hodpri,nagasawa}.
Depending on
the values of the couplings $\lambda$, $\lambda'$ and  the mass scale $M$,
we can have very different situations, indeed even when considering the same
regime the authors cited above are not always in agreement. Still,  a
reasonably definite picture emerges provided that the values of the parameters
are not too extreme (discounting the necessarily small $\lambda'$ of course),
which we now summarise before mentioning more exotic possibities.

For a rough description of what is going on, we can ignore the spatial
gradient of $\psi$, and treat it classically. As long as it is small in the
sense that
\begin{equation}
|\psi|\ll\psi_{\rm vac}
\, ,
\label{psismall}
\end{equation}
its equation of motion is
\begin{equation}
\ddot\psi + 3 H \dot \psi + M_\psi^2(\phi)\psi = 0 \,,
\label{psieq}
\end{equation}
with
\begin{equation}
M_\psi^2(\phi)=\lambda'(\phi^2-\phi^2_{{\rm inst}}) \,.
\label{mpsi}
\end{equation}
As long as $\phi>\phi_{\rm inst}$, the effective mass $M_\psi^2$ is positive
and $\psi$ is equal to zero apart from its quantum fluctuation. When $\phi$
first falls below $\phi_{\rm inst}$, $|M_\psi|$ is negligible compared with
$H$, and $\phi$ remains almost constant. After some time, which may be either
small or large on the Hubble scale depending on the regime of parameter space,
$|M_\psi|$ grows to exceed $H$, and one can start to use the opposite
approximation of ignoring $H$
\begin{equation}
\ddot \psi + M_\psi^2(\phi)\psi=0 \,.
\end{equation}
There are now two possibilities, according to whether or not the adiabatic
condition $|\dot{M}_\psi| \ll|M_\psi|^2$ is satisfied. If it is, the solution
of Eq.~(\ref{psieq}) is
\begin{equation}
\psi \simeq\mbox{constant} \times |M_\psi|^{-1/2} \exp\left(\int^t_0
|M_\psi(t)|
dt\right) \,,
\label{after}
\end{equation}
Taking $t=0$ to be the epoch when $|M_\psi|=H$,
the exponential becomes large within
a Hubble time, and Eq.~(\ref{psismall}) will be violated more or less
independently of the initial value of $\psi$. When that happens $\psi$ will
quickly roll down to its minimum $\psi_{\rm vac}$. If on the other hand the
adiabatic condition is not satisfied when $|M_\psi|$ first grows to be of
order $H$, there will typically be little change in $\psi$ until it is
satisfied, after which Eq.~(\ref{after}) will again hold. Thus the conclusion
is that $\psi$ rolls down rapidly towards its vacuum value at the epoch
$|M_\psi|\sim H$ or the epoch $|\dot{M}_\psi|/|M_\psi|^2= 1$, whichever is
later. (The insufficiency of just the former condition was pointed out in
\cite{nagasawa}.)

Though the spatial gradient of $\psi$ is not crucial initially,
it becomes so after roll down, because domain walls form
at the places in space where $\psi$ is trapped with its false vacuum value
$\psi=0$. As we already noted, more general defects can form if
$\psi$ is replaced by an $N$-component object. To determine the
stochastic properties of the spatial distribution of the defects,
one needs to consider the spatial variation of $\psi$.
The basic assumption is that during inflation $\psi$ vanishes
except for its quantum fluctuation. Once $\phi$ falls below $\phi_{{\rm
inst}}$, the fluctuation in $\psi$ can be easily evaluated, because its
Fourier modes decouple, until its {\it rms} has grown to be of order
$\psi_{\rm vac}$. The classical equation for each mode is the generalisation
of Eq.~(\ref{psieq}) including the spatial gradient
\begin{equation}
\ddot\psi_k + 3 H \dot \psi_k + \left[ \left(\frac{k}{a} \right)^2 +
	M_\psi^2(\phi) \right] \psi_k= 0 \,,
\label{psieqgrad}
\end{equation}
where $k$ is the comoving wavenumber of the mode under examination. Now
cosmologically interesting (and smaller) comoving scales presumably leave the
horizon many Hubble times after inflation begins, with the corresponding
Fourier modes initially in the vacuum, i.~e.~containing no $\psi$ particles
(\cite{LL2}, page 46). As a result the initial value of the quantum
expectation $\langle |\psi_k|^2 \rangle$ is known, and so is its time
dependence which is given simply by the modulus squared of the  solution of
the classical field equation. For the nonrelativistic modes $k/a\ll
|M_\psi|$, Eq.~(\ref{psieqgrad}) reduces to Eq.~(\ref{psieq}), and $\langle
|\psi_k|^2 \rangle$ begins to grow as the solution Eq.~(\ref{after}) becomes
valid. At this point, but not earlier, $\psi_k$ can be regarded as a classical
quantity in the sense that its quantum state corresponds to a superposition
of states with almost well defined values $\psi_k(t)$ \cite{guthpi}. Thus
after smearing the field over a distance $1/M_\psi$ (i.~e.~dropping the
relativistic modes),\footnote{Since $|M_\psi|\gg H$, these modes have yet to
leave the horizon and so are still in the vacuum state.} one has a classical
field $\psi(\bf x)$ which has a Gaussian inhomogeneity whose stochastic
properties are specified entirely by the spectrum $\langle |\psi_k|^2
\rangle$. Once the behaviour Eq.~(\ref{after}) sets in the spectrum grows
rapidly, and  the {\it rms}  of the smeared field rolls down to $\psi_{\rm
vac}$ in accordance with the earlier conclusion.

In order for this simple picture to be self consistent, the smeared field
$\psi$ must still be small enough to satisfy Eq.~(\ref{psismall}), at the
epoch when the non-relativistic modes $\psi_k$ begin to grow according to
Eq.~(\ref{after}). This can fail to be true if the parameters are far from
their natural values, for instance if $\lambda$ is very small, and then one
has a more complicated situation which has been looked at by various authors
\cite{kofpog,kbhp,sbb,hodpri,nagasawa}. In particular, the inhomogeneity in
$\psi$ might generate an adiabatic density perturbation on scales far outside
the horizon, which would then survive the subsequent phase
transition.\footnote{As has already been pointed out
\cite{linlyt,lyth92,lytste} in the somewhat different context of axions,
failure to take proper account of the phase transition has led some authors
to draw incorrect conclusions from this type of calculation. }

\subsubsection*{Topological defect production in the inflaton dominated case}

{}From the stochastic properties of $\psi(\bf x)$ just before roll down, one
can
in principle calculate the stochastic properties of the initial configuration
of the defects, since they will form at the places in space where $\psi({\bf
x})=0$.
In particular one can estimate the typical spacing of the defects.
The smallest possible spacing corresponds to the
defect size\footnote{The thickness of a domain wall or string, or the radius
of a monopole, outside which $\psi$ has its vacuum value.}, which at least for
couplings of order unity is of order $M^{-1}$. For a thermal phase transition
the
typical spacing at formation is $(\lambda M)^{-1}$ \cite{twb76}.
We would like to know if the same
is true in the inflationary case. The different estimates
\cite{vishniac,yokoyama,hodpri,nagasawa} do not entirely agree but they do
seem to indicate that the spacing is still very roughly $M^{-1}$, at least to
within a few orders of magnitude.

It does not, however, follow that the cosmological effects of the defects are
the same in the two cases, their subsequent evolution being quite
different. In the thermal case, where the defects are created during a
non-inflationary (typically radiation dominated) era, the Hubble distance
$H^{-1}$ increases steadily in comoving distance units. Except
in the case of gauge monopoles,
the
spatial distribution of the defects typically loses all memory of the initial
conditions on scales smaller than the Hubble distance, exhibiting
scaling behaviour whereby the stochastic properties
become more or less fixed in units of the Hubble distance. In particular the
typical spacing becomes of order the Hubble radius $H^{-1}$. Only on
scales much larger than $H^{-1}$ does the initial distribution expand with the
universe, remaining fixed in comoving distance units. (The case of gauge
monopoles, which do not scale, is considered in greater detail
in Section~3.3.)

The case where the defects are created during inflation is quite
different. During inflation, $H^{-1}$ decreases, typically dramatically,
in comoving distance units. As a result the
distribution of the defects is frozen in comoving distance units, and in
particular the typical spacing remains roughly of order the comoving distance
scale which left the horizon (became bigger than $H^{-1}$) at the epoch when
the defects form. This remains true until the era, long after inflation ends,
when that scale re-enters the horizon. Only then will the defect distribution
become the same as in the thermal case, as the `scaling' solution is
established.

The cosmological significance of this different evolution depends on when the
defects form. If they form {\it after} cosmologically interesting scales leave
the horizon (50 or 60 Hubble times before the end of inflation), the scaling
solution has been established by the time that these scales enter the horizon
and there should be no significant difference from the thermal case. If they
form {\it before}, their typical spacing is still much bigger than the horizon
size and we presumably see no defects (unless of course we are in an atypical
region of the universe \cite{vishniac,hodpri}). Finally, if they form at {\it
about the same time}, the configuration of the defects will differ from the
scaling solution,  as has been discussed at some length in the case of
structure forming gauge strings \cite{vishniac,yokoyama,hodpri}.

\subsection{The vacuum dominated regime}

Coming now to the regime of vacuum domination, Linde \cite{LIN2SC,LIN2SC2} has
argued that at least for couplings of order unity inflation will end promptly
(within less than a Hubble time) after $\phi=\phi_{\rm inst}$. He
demonstrated this by assuming
that $\phi$ continues to slow-roll down the potential
Eq.~(\ref{PHIPOT}) for a Hubble time, and showing that this inevitably leads
to a
contradiction. We present below
a more detailed version of this argument, repairing
an omission in the original and examining also the case of small couplings.

To proceed, one has to make the technical assumption that the spatial
gradients of both $\phi$ and $\psi$ are negligible. It is hard to see how they
could be crucial in maintaining slow-roll, so the contradiction which we
shall establish presumably indicates failure of slow-roll rather than
significant gradients in the presence of slow-roll.

The slow-roll expression for $\phi$ is
\begin{equation}
\dot\phi = - \frac{V'}{3H}=-\frac{2}{\sqrt{3\lambda}}\frac{\mpl}{\rep}
	\frac{m^2}{M^2} \phi \,,
\label{phifirst}
\end{equation}
where we have used the vacuum dominated value for $H$,
\begin{equation}
H^2 = \frac{2\pi}{3\mpl^2} \, \lambda M^4 \,.
\label{haitch}
\end{equation}
During one Hubble time the change in $\phi$ is given by
\begin{equation}
\frac{\Delta\phi}{\phi}=\frac{\dot\phi}{H \phi}=\frac{-4m^2}{\lambda
M^4} \mps \ll 1
\,.
\end{equation}
It follows that after one Hubble time
\begin{equation}
|M_\psi^2|=\frac{8m^2}{\lambda M^4} \lambda M^2\mps\ll M^2
\,,
\end{equation}
and
\begin{equation}
\frac{|M_\psi|^2}{H^2}=\frac{96m^2}{\lambda M^6} \left( \mps \right)^2
\label{mbyhs}
\,.
\end{equation}
Using Eqs.~(\ref{small}) and (\ref{large}), we see that this
is much bigger
than unity unless we are in the regime of Eq.~(\ref{small}), which then
requires
\begin{equation}
\lambda'\lsim 10^{-9} \mpsinv M^2 \lsim 10^{-9}
\,.
\label{lamprlim}
\end{equation}
We shall not consider these very small values of $\lambda'$.
One also finds after one Hubble time
\begin{equation}
\frac{|\dot{M}_\psi|\,}{|M_\psi|^2}=
	\frac12 \left(\frac{\lambda M^6}{96 m^2} \right)^{1/2} \mpsinv \,.
\end{equation}
Comparing with Eq.~(\ref{mbyhs}) one sees that except for the factor $1/2$ the
right hand side is just $H/|M_\psi|$. Thus the adiabaticity condition is
satisfied as well (it was not considered by Linde). As a result
Eq.~(\ref{after}) shows that $\psi$ will have rolled down to become of order
$\psi_{\rm vac}$ given by Eq.~(\ref{psivac}).

The next step is to demonstrate that the roll-down of $\psi$ is actually
inconsistent with slow-roll inflation. Since only one Hubble time has elapsed
$\psi$ will be oscillating around $\psi_{\rm vac}$ rather than sitting in it,
but for a crude estimate we can ignore the oscillation, so that the potential
is given by Eq.~(\ref{vnew}). Using Eqs.~(\ref{vpnew}) and (\ref{vppnew})
we find
\begin{eqnarray}
\epsilon &\simeq&
128 \frac{\lambda'}{\lambda^{3}}\frac{m^4}{M^{10}}
\left( \mps \right)^{3} \,, \\
\eta&\simeq&-\frac{8\lambda'}{\lambda M^2} \mps \,.
\end{eqnarray}
For couplings of order unity it is easy to check that the cmb constraint
Eqs.~(\ref{small}) and (\ref{large}) implies that $\epsilon$ and $|\eta|$ are
both $\gg1$. The slow-roll solution we started with is therefore presumably
invalid. (Even if valid,
it is certainly not inflationary since $\epsilon \gg 1$). More
generally, it follows from Eq.~(\ref{Mabs}) that
$\eta>1$ unless $\lambda'\lsim 10^{-4} \lambda^{1/2}$, which
again presumably means that the slow-roll solution is invalid. (The condition
that $\epsilon>1$ is too complicated to be worth discussing in the general
case, but one expects slow-roll only if both $\eta$ and $\epsilon$ are less
than unity.)

The conclusion is that (unless $\lambda'$ is very small) slow-roll inflation
ends within a Hubble time of the epoch $\phi=\phi_{\rm inst}$. Since the field
equations contain a mass scale $M\gg H$, it is reasonable to suppose that in
fact inflation ends altogether, giving way to an epoch when the energy density
is dominated by the spacetime gradients of the fields.
What happens next is associated with the question of defect production,
to which we now turn.

\subsubsection*{Defect production in the vacuum dominated case}

In contrast with the inflaton dominated case, defect production has not
previously been considered for the vacuum dominated regime. The following
discussion assumes that the couplings are of order unity, or to be more
precise that they are not extremely small for in that case a qualitatively
different scenario could ensue. In particular, we assume that $\lambda'$ is
not small enough to satisfy Eq.~(\ref{lamprlim}), so that as argued above the
phase transition marks the end of inflation.

Ideally one would like to follow the evolution of the fields using quantum
field theory in curved spacetime, as in the inflaton dominated case, and hence
calculate explicitly the typical spacing and other stochastic properties of
the initial distribution of the defects. However, that calculation relies
crucially on the fact that the back-reaction of $\psi$ on $\phi$ is negligible
which one easily checks is not the case in the vacuum dominated
regime\footnote{It is important in this connection not to be misled by the
evolution discussed in the previous subsection, which {\it was} similar to
that seen in the inflaton dominated case. That discussion was a purely
hypothetical one, used to establish a contradiction; the premise that
slow-roll continues for one Hubble time leads to that evolution, which in turn
leads to the contradictory result that slow-roll does {\it not} continue for
one Hubble time. One does not expect anything like it to actually occur, and
in particular there is no  reason to suppose that $\psi$ first rolls down to
$\psi_{\rm vac}$, after which $\phi$ falls down due to the destabilizing
action of $\psi$. (This `waterfall' sequence  of events was suggested in
\cite{LIN2SC2}.)}, and indeed an estimate using the techniques described in
Section \ref{secphidom} indicates rather that the back-reaction hits $\phi$
long before $\psi$ has had a chance to roll down. In the absence of this
simplifying feature, it is not even possible to give a qualitative account of
the evolution, let alone follow it in detail. When $\phi$ is hit by the
back-reaction from $\psi$, it will acquire a spatial gradient of order
$M_\psi$, which will soon become much bigger than $m$. As a result of
the backreaction, $\phi$
will look more like a collection of interacting plane waves than a homogeneous
field, and cannot be said to `roll down'
to its true vacuum. Moreover, $\psi$ will in general still be an essentially
quantum object at this stage (i.~e.~the state of the system is not a
superposition of states in which it has an almost well defined value over an
extended period of time), so $\phi$ will become one as well.

In the absence of an explicit calculation one must rely on order of magnitude
arguments, which as we now see actually point to rather definite conclusions.
The crucial point is that the $\phi$ and $\psi$ fields are coupled to each
other, and also in general to the quark, lepton etc.~fields. Since at least
the scale $M$ is much bigger than $H$ (and even $m$ is not many orders of
magnitude less), one expects the fields to thermalise quickly on the
Hubble timescale; reheating in the vacuum dominated case will occur
promptly at the end of inflation. The reheat temperature $T_{{\rm reh}}$ is
therefore given by the familiar formula $T_{{\rm reh}}=(30/\pi^2 g_*)^{1/4}
\rho^{1/4}$, or
\begin{equation}
T_{{\rm reh}}=(30\lambda/4\pi^2 g_*)^{1/4} M \,,
\label{treh}
\end{equation}
where $g_*$ is the effective number of degrees of freedom at that
temperature, presumably at least of order $10^2$ (e.~g.~in the minimal
supersymmetric standard model $g_*=229$). Thus the reheat temperature is of
order $M$.
The defects that have been formed find themselves effectively in a
thermal bath at a temperature $T_{\rm reh}$,
and may be in thermal equilibrium. If so, then for a string network, for
example,
the most likely configuration
will be the one
which maximises the allowed density of states.
For the case of cosmic strings, such a configuration,
below the Hagedorn transition consists of maximising the
number of possible loops that can form, with the long
strings being
exponentially suppressed \cite{copeland90}. Since
this distribution is similiar to that found soon after
a thermal phase transition has produced strings, it is possible that
the effect of such a
rapid reheating could lead to a configuration of defects
much the same as if there had been a thermal phase transition. However,
we have not explicitly
demonstrated this to be the case here; it would require a detailed
numerical simulation of the reheating to rigorously establish how the
network behaves.

Local cosmic strings are perhaps the most interesting defect for cosmology.
The primary motivation for employing them here
would be so they could contribute as seeds for the observed
large scale structure and the anisotropies in the microwave background.
However,
if we do try to make use of them in the context of this inflation
model, we must be cautious; it is expected that strings would have a very
important influence on the cmb if they are to be massive enough to
affect structure formation \cite{Albert}.
Therefore we must reassess the
estimates of the allowed model parameters determined in the
previous section, for these were obtained assuming that
the inflaton field alone was responsible for the cmb anisotropies.
With two sources (assumed uncorrelated) of anisotropies, the contributions
add in quadrature. As a benchmark figure we reduce the inflation contribution
to 10\% of the total anisotropy, corresponding to dropping $\delta_H$ by
$\sqrt{10}$. Numerical calculation shows that for unit couplings, this only
reduces $M_{max}$ from $2.4 \times 10^{-3} \mpl$ to $1.3 \times 10^{-3}
\mpl$.

Recent simulations of cosmic string networks has shown the importance of
the small scale structure on the network. An important effect of this structure
is to renormalise the string mass per unit length $\mu$, relating it to the
original mass per unit length $\mu_0$ by $\mu \sim 1.4 \mu_0$ \cite{tanmay}.
Recalling that $\mu_0 \simeq \pi M^2$ (provided scalar and vector masses are
not too disparate), we are therefore
allowed a mass per unit length of
up to $9 \times 10^{-6} \mpl$, which is
comfortably high enough to allow $\mu$ to fall in the favoured range
of values
for structure formation $\mu \sim 2-4 \times 10^{-6} \mpl^2$ \cite{AS92}.
An equivalent
calculation indicates that global textures can be similarly reconciled, though
more marginally.  Indeed, for the
highest values we can obtain, the strings would create excessive microwave
fluctuations; the best current bound on $\mu$ arises
from cmb anisotropies on scales less than 10', and yields
$\mu < 3 \times 10^{-6} \mpl^2$ \cite{hindmarsh93}. Defect production
can therefore provide an additional constraint on the viable parameters of the
inflationary theory.

Finally,
recall that the above is with the unfavourable assumption of unit couplings;
the analysis of Section \ref{INFL} indicates that the upper limit
on $M$ is yet higher if the couplings are reduced, which to lowest order does
not alter the string tension.

\subsection{Non-thermal monopole production}
\label{monopole}

We now consider gauge
monopole production in a non-thermal phase transition.
Some aspects of such production have already been disussed in \cite{Laycock}.
The case of
gauge monopoles is particularly interesting because they do not reach
a scaling regime. Let the initial correlation length be some fraction $\zeta$
of the Hubble radius, $\xi \equiv \zeta H^{-1}$. For the thermal case, it is
easy to show that $\zeta_{\rm th} \sim g_*^{1/2}M/ \lambda \mpl$. In the
vacuum dominated case all we can be sure of (for fast roll-down) is that
$\zeta_{{\rm vac}} \le 1$. Of course the uncertainties in the initial
distribution will be reflected in our lack of knowledge of the form
$\zeta_{{\rm vac}}$ should take.

Now in general we can write the initial number density of monopoles
and temperature as
\begin{equation}
{n_{\rm i}\over T_i^3} \sim {1\over \xi^3 T_i^3} \sim {1\over \zeta^3}
{H^3 \over T_i^3}.
\label{ninitial}
\end{equation}
Once we know $T$ we can determine $H$, and hence the future evolution
of $n/T^3$ for a given value of $\zeta$.
We have demonstrated that in the false vacuum dominated
case, reheating is prompt, leading to a temperature after inflation given by
Eq.~(\ref{treh}). Now this means that in Eq.~(\ref{ninitial}), for a given
reheat
temperature and assuming $\zeta \leq 1$,
we obtain a similar scenario to the thermal case \cite{rockybook}.
Neglecting the effects of annihilation, which should be valid for
monopole masses,
$M_{\rm mon} \le 10^{17} {\rm GeV}$ and $\zeta \sim 1$, we obtain
\begin{equation}
\Omega_{\rm mon} h^2 \simeq {10^{11} \over \zeta^3} \left({T_{\rm reh}
\over 10^{14}~{\rm GeV}}\right)^3 \left({M_{\rm mon} \over
10^{16}~{\rm GeV}}\right),
\end{equation}
where $h$ is the Hubble parameter today in units of
$100 {\rm kms}^{-1} {\rm Mpc}^{-1}$ and $0.4 < h < 1$ \cite{rockybook}.
In other words, we are unable to differentiate between the thermal
production of monopoles and this particular non-thermal case, for a
given initial temperature. The original GUT scale
monopole problem still exists in
the non-thermal case.

Demanding $\Omega_{\rm
mon}h^2 < 1$ constrains us to a region $M_{\rm mon} < 10^{13} {\rm GeV}$.
This is not the strongest bound though, especially for the case
of light monopoles. The Parker limit \cite{parker}
(see \cite{rockybook} for
details), places a constraint on the allowed flux in
monopoles of mass below $10^{17}$GeV. For consistency we find that $
M_{\rm mon} < 10^{12}$ GeV, slightly tighter than the density bound.
Thus it appears that at least under the assumption that $\zeta \sim 1$,
the monopole problem is still very much present in the vacuum
dominated region.

\section{Particle Physics Models}
\label{SUGR}
\setcounter{equation}{0}
\def\theequation{\thesection.\arabic{equation}}

No matter how simple it might be, and no matter how well its predictions
agree with observation, no model of inflation can be regarded as
satisfactory unless it emerges from a sensible theory of particle
physics. In the present context (Eq.~(\ref{FULLPOT})) this means that
we want to identify $\phi$ and $\psi$ with fields belonging to such
a theory, {\it and} to show that $\phi$ can have a sufficiently flat
potential without fine tuning, in particular to show that the mass
$m$ of the $\phi$ field can be sufficiently small ($m \ll H$) and
that there is no $\phi^4$ term.

We start by considering the case of global supersymmetry \cite{Nilles}.
Here it is natural to focus on the regime $100\,$GeV to 1\,TeV for $m$,
which is the smallest one commonly considered for scalar fields. The
requirement of having no $\phi^4$ term means that one cannot identify
$\phi$ with a Higgs field, but it might be one of the scalar fields
suggested by supersymmetric theories. With $m$ in this range, the COBE
normalisation requires that $M$ is of order $10^{11}\,$GeV which,
as noted earlier \cite{LL2}, suggests the possibility that the false
vacuum is that of Peccei-Quinn symmetry.
One is therefore led to ask whether, by considering Peccei-Quinn symmetry
in the context of supersymmetry, there emerges a field $\phi$ with a
quadratic coupling to the Peccei-Quinn field and a mass of order
$1\,\mbox{\rm TeV}$, but {\it no $\phi^4$ coupling}.
For global supersymmetry the answer to our question is remarkable; the
very first model of supersymmetric Peccei-Quinn symmetry, proposed by Kim in
1984 \cite{Kim}, indeed has a suitable field $\phi$. If we could stop at this
point, we would have fulfilled the wildest dreams of particle cosmologists.
A model motivated purely by particle physics would subsequently
be seen (in this case nine years later) to lead to an observationally
viable epoch of inflation without any fine tuning of its parameters!

Unfortunately, there are sound particle physics reasons for rejecting global
supersymmetry and replacing it by supergravity \cite{Nilles}.
In a general supergravity theory it is difficult to construct a model of
inflation without fine tuning, because in addition to the
global supersymmetric type terms there is an infinite series of higher order
non-renormalisable terms, the first of which usually gives any would-be
inflaton an effective mass of order $H$.

However, supergravity can only be regarded as an effective theory with
a cutoff at the Planck scale. So in order to get a better handle on the
crucial non-renormalisable terms we should consider a theory of everything.
Superstrings provide one possible candidate theory. Here we find
another superstring miracle. For a class of
low energy effective supergravity theories
derived from superstrings, they are of precisely
the form necessary to cancel the
harmful non-renormalisable terms. We go on to make the first steps towards
constructing a truely realistic, superstring derived, model of inflation.

\subsection{A simple supersymmetric model}
\label{sa}

The simplest superpotential \cite{Nilles} that spontaneously breaks a
U(1) symmetry is
\begin{equation}
\label{sa1}
W = \sigma ( \Psi_1 \Psi_2 + \Lambda^2 ) \Phi \,,
\end{equation}
where $\Phi$, $\Psi_1$ and $\Psi_2$ are chiral superfields
which we take to have canonical kinetic terms, $\Lambda$ is
a mass which sets the scale of the spontaneous symmetry breaking,
$\sigma$ is a coupling constant, and the U(1) symmetry is
$\Psi_1 \rightarrow e^{i\theta} \Psi_1$,
$\Psi_2 \rightarrow e^{-i\theta} \Psi_2$.
This superpotential is often used in supersymmetric model
building \cite{Nilles,Kim},  and in particular was used by Kim \cite{Kim}
to construct the first supersymmetric realisation of Peccei-Quinn symmetry.
We shall now show that for fairly generic initial conditions, it leads to
the false vacuum inflation model of the previous sections with the
identifications $\lambda'= 2\lambda = \sigma^2/2$ and $M = 2\Lambda$.

The scalar potential derived from this superpotential is
\begin{eqnarray}
V & = & \left| \frac{\partial W}{\partial \Phi} \right|^2
	+ \left| \frac{\partial W}{\partial \Psi_1} \right|^2
	+ \left| \frac{\partial W}{\partial \Psi_2} \right|^2 \,, \\
  & = & \sigma^2 \left| \Psi_1 \Psi_2 + \Lambda^2 \right|^2
    + \sigma^2 \left( |\Psi_1|^2 + |\Psi_2|^2 \right) |\Phi|^2 \,,
\end{eqnarray}
where $\Phi$, $\Psi_1$ and $\Psi_2$ now represent just the (complex)
scalar component of the respective chiral superfields.
Adding a soft supersymmetry breaking mass, $m$, of order 1 TeV, for
$\Phi$, we obtain
\begin{equation}
\label{sa2}
V =  \sigma^2 \left| \Psi_1 \Psi_2 + \Lambda^2 \right|^2
    + \sigma^2 \left( |\Psi_1|^2 + |\Psi_2|^2 \right) |\Phi|^2
	+ m^2 |\Phi|^2 \,.
\end{equation}

We want to show that $\Phi$ can be the inflaton, and so to obtain the
effective potential during inflation we will minimise this potential for
fixed $\Phi$.
The potential is minimised at $\arg \Psi_1 + \arg \Psi_2 = \pi$, and the
canonically normalised field corresponding to the phase of the $\Psi$ field
with smaller magnitude has an effective mass $\geq \sigma \Lambda$ there.
At least if $\sigma$ is not very small this will anchor
$\arg \Psi_1 + \arg \Psi_2$ at the value $\pi$.
The potential is independent of the other two angular degrees of freedom,
namely $\arg \Psi_1 - \arg \Psi_2 $ (which corresponds to the axion field)
and $\arg\Phi$, and Hubble damping will make them practically time
independent more or less independently of the initial conditions.
This leaves only the radial degrees of freedom,
corresponding to the three canonically normalised real fields
$\phi = \sqrt{2}\,|\Phi|$, $\psi_1 = \sqrt{2}\,|\Psi_1|$ and
$\psi_2 = \sqrt{2}\,|\Psi_2|$. In terms of them, the potential is
\begin{equation}
V(\phi , \psi_1 , \psi_2 ) =
    \frac{\sigma^2}{4} \left( \psi_1 \psi_2 - 2\Lambda^2 \right)^2
    + \frac{\sigma^2}{4} \left(\psi_{1}^{2}+\psi_{2}^{2}\right)\phi^2
	+ \frac{1}{2} m^2 \phi^2 \,.
\end{equation}
For $\phi > 0$, the degree of freedom orthogonal to $\psi_1 \psi_2$
(which corresponds to the saxino) has its minimum at $\psi_1 = \psi_2$,
and it is straightforward to show that the effective mass of the canonically
normalised saxino field is everywhere $\geq \sigma \phi / \sqrt{2}$ and so
the saxino will be firmly fixed at its minimum during inflation\footnote{
$\phi > \phi_{\rm inst} = \sqrt{2}\, \Lambda$ during inflation
(see Eq.~(\ref{sa9}) and Sections~2 and~3).}.
Then in terms of $\phi$ and the canonically normalised field
$\psi = \sqrt{2 \psi_1 \psi_2}$, the potential is
\begin{equation}
\label{sa9}
V(\phi,\psi) = \frac{\sigma^2}{16} \left(\psi^2 - 4\Lambda^2 \right)^2
    + \frac{\sigma^2}{4} \phi^2 \psi^2 + \frac{1}{2} m^2 \phi^2 \,.
\end{equation}
Thus we have the model of Sections~2 and 3, with $\lambda'=2\lambda
=\sigma^2/2$ and $M=2\Lambda$. From Eq.~(\ref{extvacdom}) it follows
that the usual axion parameter $f_a$ is given by
\begin{equation}
\label{sa10}
f_{a} = 2\Lambda = 8\times 10^{11}\mbox{GeV}
\;\sigma^{-\frac{3}{5}}\left(\frac{m}{\mbox{TeV}}\right)^{\frac{2}{5}} \,,
\end{equation}
which is at the right scale for the axion.

Note that for chaotic initial conditions $\Lambda$ and $m$ will initially
be negligible and so the potential Eq.~(\ref{sa2}) will initially have
the simple form
\begin{equation}
V = \sigma^2 \left( \left| \Psi_1 \Psi_2 \right|^2
	+ \left| \Psi_2 \Phi \right|^2 + \left| \Phi \Psi_1 \right|^2 \right)
\,.
\end{equation}
Thus if initially $|\Phi| > |\Psi_1|, |\Psi_2|$, {\mbox i.e.} one third
of the initial condition space, the fields will rapidly approach the
inflating trajectory $|\Psi_1| = |\Psi_2| = 0$ and $|\Phi| \gg \Lambda$
given above.

\subsection{Supergravity}
\label{sg}

The scalar potential in supergravity \cite{Nilles} has the general form
\begin{eqnarray}
V & = & \exp \left( \frac{8\pi}{\mpl^2} K \right) \left[ \sum_{\alpha,\beta}
\left( \frac{\partial^2 K}{\partial \bar{\phi}_{\alpha} \partial \phi_{\beta} }
	\right)^{-1}
\left( \frac{\partial W}{\partial \phi_{\alpha} } + \frac{8\pi}{\mpl^2}
	W \frac{\partial K}{\partial \phi_{\alpha} } \right)
\left( \frac{\partial \bar{W} }{\partial \bar{\phi}_{\beta} }
	+ \frac{8\pi}{\mpl^2} \bar{W}
	\frac{\partial K}{\partial \bar{\phi}_{\beta} } \right)
- 3 \frac{8\pi}{\mpl^2} |W|^2 \right] \nonumber \\
\label{sg1}
 & & + {\rm \ D-term} \,,
\end{eqnarray}
where the K\"{a}hler potential $K( \phi , \bar{\phi} )$ is a real
function of the complex scalar fields $\phi_{\alpha}$ and their hermitian
conjugates $\bar{\phi}_{\alpha}$, and the superpotential $W(\phi)$ is an
analytic function of $\phi$. The D-term is quartic in the charged fields,
and we will assume that it is flat along the inflationary trajectory so that
it can be ignored during inflation. It may however play a vital role in
determining the trajectory and in stabilising the non-inflaton fields.
The term given explicitly is called the F-term.

The kinetic terms are
\begin{equation}
\label{sg15}
\sum_{\alpha,\beta}
\frac{\partial^2 K}{\partial \phi_{\alpha} \partial \bar{\phi}_{\beta}}
\partial_\mu  \phi_{\alpha} \partial^\mu \bar{\phi}_{\beta} \,,
\end{equation}
where $\mu$ is a spacetime index. It follows that
for canonically normalised fields
\begin{equation}
\label{sg2}
K = \sum_{\alpha} \left| \phi_{\alpha} \right|^2 + \ldots \,,
\end{equation}
where \ldots\ stand for higher order terms.
Global supersymmetry corresponds to the case where these terms are
absent, and one has taken the limit $\mpl\to\infty$ to obtain the
potential $V=\sum_\alpha|\partial W/\partial \phi_\alpha|^2$ that
we used earlier. Supergravity corresponds to keeping $\mpl$ finite.
Then the F-term part of the scalar potential becomes
\begin{eqnarray}
\label{sg3}
V & = & \exp \left( \frac{8\pi}{\mpl^2} \sum_{\gamma}
	\left| \phi_{\gamma} \right|^2 + \ldots \right) \times \\
 & &
\left\{ \sum_{\alpha,\beta} \left( \delta_{\alpha \beta} + \ldots \right)
\left[ \frac{\partial W}{\partial \phi_{\alpha} } + \frac{8\pi}{\mpl^2}
	\left( \bar{\phi}_{\alpha} + \ldots \right) W \right]
\left[ \frac{\partial \bar{W} }{\partial \bar{\phi}_{\beta} }
	+ \frac{8\pi}{\mpl^2} \left( \phi_{\beta} + \ldots \right) \bar{W}
\right]
- 3 \frac{8\pi}{\mpl^2} |W|^2 \right\} \,. \nonumber
\end{eqnarray}
Thus, for {\em any} model of inflation, the lowest order
({\mbox i.e.} global supersymmetric) inflationary potential
\begin{equation}
V_{\rm global} \equiv \sum_{\alpha}
	\left| \frac{\partial W}{\partial \phi_{\alpha} } \right|^2 \,,
\end{equation}
will receive corrections\footnote{
We assume that the corrections are small as will be the case if
$ \left| \phi_{\alpha} \right| \ll \mpl/\sqrt{8\pi} $. If
$ \left| \phi_{\alpha} \right| \gsim \mpl/\sqrt{8\pi} $ then a glance
at the exponential factor in Eq.~(\ref{sg3}) shows that the problems
will then be even more severe.}
giving
\begin{equation}
\label{sg4}
V = V_{\rm global} \left( 1 + \frac{8\pi}{\mpl^2} \sum_{\alpha}
	\left| \phi_{\alpha} \right|^2 + {\rm other\ terms} \right)
+ {\rm other\ terms} \,.
\end{equation}
The $ \left| \phi_{\alpha} \right|^2 $ term in this equation gives a
contribution $ 8\pi V_{\rm global} / \mpl^2 \simeq 3H^2 $ to the effective
mass squared of {\em all} scalar fields, therefore, assuming the inflaton is
the modulus of a scalar field\footnote{ Natural inflation \cite{natural}
avoids this problem because its inflaton is the phase of a complex scalar
field.}, it gives a contribution of order unity
to $\eta \equiv \mpl^2 V''/8\pi V$
(see Section~\ref{infldyn}). But $|\eta| \ll 1$ is necessary for
inflation to work (at least in the usual slow-roll form).
As a result practically all\footnote{ The only exception known to us is
`natural inflation' \cite{natural}, as mentioned above.} of the supergravity
models of inflation proposed so far \cite{olive} have involved unmotivated
fine tuning of the K\"ahler potential and/or the superpotential in order to
cancel the harmful non-renormalisable corrections ({\mbox i.e.} to get the
`other terms' in Eq.~(\ref{sg4}) to cancel the
$ \left| \phi_{\alpha} \right|^2 $ term).
However, supergravity can only be regarded as an effective theory with
a cutoff at the Planck scale. So in order to get a better handle on the
crucial non-renormalisable terms we should consider a theory of everything.
Superstrings provide the most promising candidate, and in the
next section we will find that for the K\"ahler potential derived from
orbifold compactification of superstrings the cancellation can occur
without any fine tuning.

To end this section we just note that for the special choice of supergravity
with minimal kinetic terms\footnote{ Although this is in some sense the
simplest supergravity theory, it is not well motivated physically and its
adoption must be regarded as fine tuning to some extent.} ({\mbox i.e.}
Eq.~(\ref{sg2}) and Eq.~(\ref{sg3}) without the higher order corrections
\ldots ) and the superpotential of the previous section, the `other terms'
in Eq.~(\ref{sg4}) cancel the $ \left| \phi_{\alpha} \right|^2 $ term for
the inflaton, as we will now show.

Substituting $ K = |\Phi|^2 + |\Psi_1|^2 + |\Psi_2|^2 $ and the superpotential
of the previous section, Eq.~(\ref{sa1}), into Eq.~(\ref{sg1}) gives
\begin{eqnarray}
V( \Phi , \Psi_1 , \Psi_2 ) & \simeq &
	\sigma^2 \exp \left( \frac{8\pi}{\mpl^2} |\Phi|^2 \right)
	\times \\
 & & \left[ \left| \Psi_1 \Psi_2 - \Lambda^2 \right|^2 \left( 1 -
	\frac{8\pi}{\mpl^2} |\Phi|^2 + \frac{(8\pi)^2}{\mpl^4} |\Phi|^4
	\right) + \left( |\Psi_1|^2 + |\Psi_2|^2 \right) |\Phi|^2 \right]
\,. \nonumber
\end{eqnarray}
Minimising with respect to $\Psi_1$ and $\Psi_2$ for $|\Phi|>\Lambda$
as in the previous section, gives
\begin{eqnarray}
V(\phi) & = & \sigma^2 \Lambda^4 \exp \left( \frac{1}{2} \frac{8\pi}{\mpl^2}
	\phi^2 \right) \left( 1 - \frac{1}{2} \frac{8\pi}{\mpl^2} \phi^2
	+ \frac{1}{4} \frac{(8\pi)^2}{\mpl^4} \phi^4 \right) \,,
	\nonumber \\
  & = & \sigma^2 \Lambda^4 \left( 1 + \frac{1}{8} \frac{(8\pi)^2}{\mpl^4}
	\phi^4 + \ldots \right) \,.
\end{eqnarray}
Thus the problematic mass term cancels out. However we do not regard this
as a realistic model and so will not pursue it further.

\subsection{Superstrings}
\label{ss}

The inflation that inflated the observable universe beyond the
Hubble radius, and could have produced the seed inhomogeneities
necessary for galaxy formation and the anisotropies recently
observed by COBE, must occur at an energy scale
$V^{1/4} \leq 4 \times 10^{16}$GeV \cite{LiddleES},
well below the Planck scale.
At these relatively low energies, superstrings are described by an
effective N=1 supergravity theory \cite{Nilles}. The properties of
that supergravity theory are known in most detail for orbifold
compactification schemes, and so we will restrict ourselves to
such compactifications, although our results may be more general.
Also, for simplicity, we will ignore the twisted sector of the
theory. For the remainder of this section we set $\mpl/\sqrt{8\pi} = 1$.

Following \cite{Lust}, we will assume the following form for the one-loop
corrected K\"{a}hler potential $K$ of the supergravity theory derived from
orbifold compactification of superstrings \cite{Ferrara,GS,Lust,Carlos}
\begin{equation}
\label{ss1}
K = - \ln Y - \sum_{i=1}^{3} \ln X_i \,,
\end{equation}
with
\begin{equation}
Y = S + \bar{S} + \frac{1}{4\pi^2} \sum_{i=1}^{3} \delta^{\rm GS}_{i}
    \ln X_i \,,
\end{equation}
and
\begin{equation}
X_i = T_i + \bar{T}_{i} - \sum_{\alpha} \left|\phi_{i}^{\alpha}\right|^2 \,,
\end{equation}
where $S$ is the dilaton whose real part gives the tree-level
gauge coupling constant (Re$\,S \sim g_{\rm gut}^{-2}$),
$T_{i}$ are untwisted moduli whose real parts give the radii of the
three compact complex dimensions of the orbifold, and
$\phi_{i}^{\alpha}$ are the untwisted matter fields associated with $T_{i}$.
The terms with coefficients $\delta^{\rm GS}_{i}$ are one-loop corrections
coming from the Green-Schwarz mechanism, whose matter field dependence
is speculative \cite{Lust} (note that our convention for the sign and
magnitude of these coefficients follows \cite{GS,Carlos}, not \cite{Lust}).

For initial orientation we make the standard assumption that the dilaton and
moduli have expectation values of order one, showing later how this may be
achieved. (In our vacuum, $\langle {\rm Re}\,S \rangle \sim 2$ and
$\langle {\rm Re}\,T_i \rangle \sim 1$ \cite{Carlos}, but as we note
later these values may be different during inflation.)
We will also make the standard assumption that the matter fields have
expectation values much less than one. The fact that this can and does
include the inflaton is an important advantage of this model of inflation.
The values of the dimensionless coefficients $\delta^{\rm GS}_i$ depend on
the orbifold assumed. Some values for $\delta^{\rm GS}_{i}$ that have been
calculated in \cite{GS} are 0, 5 and 15. We will assume
$\delta^{\rm GS}_{i} \geq 0$ as is the case for all orbifolds considered up
to now \cite{GS,Brignole}.
% and in particular will concentrate on a class of
%models in which the one associated with the inflaton is nonzero,
%and solely responsible for the slope of the inflationary potential
%and therefore for the tilt $n-1$ of the spectrum.

The F-term part of the scalar potential corresponding to this
K\"{a}hler potential is \cite{Lust}
{\samepage
\begin{eqnarray}
\label{ss2}
V & = & \frac{1}{ Y \prod_{i=1}^{3} X_i } \left\{
\left| W - Y \frac{\partial W}{\partial S} \right|^2 - 3|W|^2
\right. \\
 & + & \left.
\sum_{i=1}^{3} \frac{Y}{Y+\frac{1}{4\pi^2}\delta^{\rm GS}_{i}}
\left( \left| W + \frac{1}{4\pi^2} \delta^{\rm GS}_{i}
\frac{\partial W}{\partial S} - X_i \frac{\partial W}{\partial T_i} \right|^2
+ X_i \sum_{\alpha} \left| \frac{\partial W}{\partial \phi_{i}^{\alpha}}
+ \bar{\phi}_{i}^{\alpha}\frac{\partial W}{\partial T_{i}} \right|^2
\right) \right\} \,. \nonumber
\end{eqnarray}
}
To lowest order in $\delta^{\rm GS}$ and the matter fields, the
kinetic terms given by Eq.~(\ref{sg15}) are
\begin{equation}
\label{ss3}
\frac{1}{ \left( S + \bar{S} \right)^2 } \partial_\mu S \partial^\mu \bar S
+ \sum_i \frac{1}{ \left( T_i + \bar{T}_i \right)^2 }
	\partial_\mu T_i \partial^\mu \bar T_i
+ \sum_{i,\alpha} \frac{1}{ T_i + \bar{T}_i } \partial_\mu \phi_{i}^{\alpha}
	\partial^\mu \bar{\phi}_{i}^{\alpha} \,.
\end{equation}
The superpotential $W$ is composed of a perturbative part,
$W_{\rm pert}(\phi,T)$, and a non-perturbative part, $W_{\rm np}(\phi,S,T)$.
To lowest order in the matter fields, $W_{\rm pert}$ has the general form
\cite{Ferrara,Lust,Carlos}
\begin{equation}
W_{\rm pert} = \sum_{\alpha,\beta,\gamma} w_{\alpha\beta\gamma}
	\phi_{1}^{\alpha} \phi_{2}^{\beta} \phi_{3}^{\gamma} \,,
\end{equation}
where $w_{\alpha\beta\gamma} = 0$ or 1.
$W_{\rm np}$ is not very well understood. However, it should have an
expansion in powers of $e^S$ to reflect its nonperturbative nature
\cite{Cvetic,Carlos}.
Also, orbifold compactifications of superstrings are invariant under
target-space duality symmetries to all orders of string perturbation theory
and, it is thought, nonperturbatively as well \cite{Theisen,Cvetic,Ibanez}.
These duality transformations act on the moduli as
\begin{equation}
T_i \rightarrow \frac{a_i T_i - i b_i}{i c_i T_i + d_i} \,, \;\;\;\;
a_i d_i - b_i c_i = 1 \,.
\end{equation}
The parameters $a_i, b_i, c_i, d_i$ are in general a discrete set of real
numbers. In many cases the duality group is given by the product of three
modular groups, {\mbox i.e.} $a_i, b_i, c_i, d_i \in {\bf Z}$, and for
simplicity we will assume this to be the case here. Then the matter fields
$\phi_{i}^{\alpha}$ transform in the same way as $ 1 / [ \eta(T_i) ]^2 $
where
\begin{equation}
\eta(T_i) = e^{-\frac{\pi T_i}{12}} \prod_{n=1}^{\infty}
\left( 1 - e^{-2n\pi T_i} \right)
\end{equation}
is the Dedekind function. It will also be useful to define the
modular-invariant dilaton field \cite{GS,Lust}
\begin{equation}
S' = S - \frac{1}{4\pi^2} \sum_{i=1}^{3} \delta^{\rm GS}_{i}
\ln \left[ \eta \left( T_i \right) \right]^2
\end{equation}
from which the transformation properties of the dilaton can be deduced.
Requiring modular invariance then puts strong constraints on the form of the
low-energy supergravity theory and in particular on $W_{\rm np}$
\cite{Theisen,Cvetic,Lust,Carlos}.

As we shall see, the K\"{a}hler potential of Eq.~(\ref{ss1}) has some very
special properties as far as inflation is concerned.
The crucial point is the cancellation of the $X_i$ factor in front of the
global supersymmetric $\partial W / \partial \phi_{i}^{\alpha}$ term in
Eq.~(\ref{ss2}).
This has the consequence that if the
$\partial W / \partial \phi_{i}^{\alpha}$ terms for one value of $i$,
say $i=3$, dominate the inflationary potential energy then the canonically
normalised $T_3$ and
$\phi^\alpha_3$ fields do {\it not} acquire corrections\footnote
{An exception to this statement is the case of a $\phi_3^\alpha$
field whose
$\partial W / \partial \phi_{3}^{\alpha}$ terms contribute to the
inflationary potential energy and are likely to pick up masses of order $H$
from the $|W|^2$ terms in Eq.~(\ref{ss2}).} of order $H$
to their effective masses as
would be expected in supergravity in general (see the previous section).
This opens up a path to inflation without fine tuning.
Note that the above conditions for inflation are asymmetric in the $T_i$
which means that the K\"{a}hler potential
$K = - \ln \left( S+\bar{S} \right) - 3\ln \left( T+\bar{T}
- \sum_{\alpha} \left|\phi^{\alpha}\right|^2 \right)$ \cite{Witten}
cannot be regarded as equivalent to the K\"{a}hler potential of Eq.~(\ref{ss1})
and does not share its inflationary properties.
We will now go on to chart this path to inflation for the case of false
vacuum inflation.\footnote{ One of us will consider the case of true
vacuum inflation elsewhere \cite{Ewan}.}

Since we are assuming $\left| \phi_{i}^{\alpha} \right| \ll 1$, it is not
unreasonable to assume that the $\partial W / \partial \phi_{i}^{\alpha}$
terms dominate the potential during inflation.\footnote{ We will consider
what effect the other terms might have on the inflaton's potential later.}
Then
\begin{equation}
\label{ss6}
V = \sum_{i=1}^{3} \frac{ \sum_{\alpha}
	\left| \frac{\partial W}{\partial \phi_{i}^{\alpha}} \right|^2 }
{ \left( Y + \frac{1}{4\pi^2}\delta^{\rm GS}_{i} \right)
\prod_{j \neq i} X_j } \,.
\end{equation}

Next we minimise the potential, with respect to the matter field dependence
of the $\partial W / \partial \phi_{i}^{\alpha}$ terms,

For a fixed
inflationary value ({\mbox i.e.} $\phi > \phi_{\rm inst}$, see Section~2)
of the inflaton (whatever it may turn out to be),
the matter fields have masses much greater than $S$, so they will settle
to their values on the inflationary trajectory before $S$ moves
significantly. We assume that the
$\partial W / \partial \phi_{i}^{\alpha}$ terms will then be independent of
the inflaton, as will be the case for false vacuum inflation.
We also assume that some of them will be non-zero.
Dropping temporarily the superscripts on the matter fields, $W$ transforms
under modular transformations like $\phi_1 \phi_2 \phi_3$, and therefore
$\partial W / \partial \phi_{i}$ transforms like $\prod_{j\neq i} \phi_j$
which we noted earlier transforms like $\prod_{j\neq i} [1/\eta(T_j)]^2$.
Since the $S$ dependence arises entirely from non-perturbative effects,
one therefore expects the following functional form
\cite{Theisen,Cvetic,Lust,Carlos}
\begin{equation}
\label{ss62}
\frac{ \partial W }{ \partial \phi_{i}^{\alpha} }
= \frac{ \sum_n a_{in}^{\alpha}(T) e^{-b_n S'} }
{ \prod_{j \neq i} \left[\eta \left(T_j \right) \right]^2 } \,,
\end{equation}
where the $a_{in}^{\alpha}$'s will in general be arbitrary modular invariant
functions of the $T_i$, but for the most part we will assume that they are
constants as is the case in gaugino condensation scenarios \cite{Carlos}.
Also, we will assume that the $b_n$'s are positive as in such scenarios.

Now, as will soon become clear, in order to get inflation we need the false
vacuum energy density, $V_0$, to be dominated by one or more
$\partial W / \partial \phi_{i}^{\alpha}$ terms with the same value of $i$,
say $i=3$. For the moment let us ignore the other $i$ values altogether.
Then the potential during inflation is given by
%%%%%%%%%%%%%%%%%%%%%%%%%%%%%%%%%%%%%%%%%%%%%%%%%%%%%%%%%%%%%
\begin{eqnarray}
\label{ss65}
V_{\rm infl} & = & \frac{ \sum_{\alpha}
	\left| \frac{\partial W}{\partial \phi_{3}^{\alpha}} \right|^2 }
{ \left( Y + \frac{1}{4\pi^2}\delta^{\rm GS}_{3} \right)
\prod_{i \neq 3} X_i } \,, \\
 & = &
\frac{ \sum_{\alpha} \left| \sum_n a_{3n}^{\alpha} e^{-b_n S'} \right|^2 }
{AB}
\end{eqnarray}
where
\begin{eqnarray}
A&\equiv&
S' + \bar{S'} + \frac{1}{4\pi^2} \delta^{\rm GS}_{3}
	+ \frac{1}{4\pi^2} \sum_{i=1}^{3} \delta^{\rm GS}_{i}
	\ln \left[ \left( T_i + \bar{T}_i
	- \sum_{\beta} \left| \phi_{i}^{\beta} \right|^2 \right)
	\left| \eta \left( T_i \right) \right|^4 \right] \nonumber\\
B&\equiv &\prod_{i \neq 3} \left( T_i + \bar{T}_i
	- \sum_{\beta} \left| \phi_{i}^{\beta} \right|^2 \right)
	\left| \eta \left( T_i \right) \right|^4  \,. \nonumber
\end{eqnarray}
This may be written in the form
\begin{equation}
\label{ss7}
V
= V_0 \left\{ 1 + \sum_{i \neq 3} \sum_{\beta}
	\frac{ \left| \phi_{i}^{\beta} \right|^2 }{ T_i + \bar{T}_i }
+ \frac{ \delta^{\rm GS}_{3} }{ 4\pi^2 \left( S + \bar{S} \right) }
\left\{
\sum_{\beta} \frac{ \left| \phi_{3}^{\beta} \right|^2 }{ T_3 + \bar{T}_3 }
	-\ln \left[ \left( T_3 + \bar{T}_3 \right)
	\left| \eta \left( T_3 \right) \right|^4 \right]\right\}
+ \ldots \right\} \,,
\end{equation}
where
\begin{equation}
V_0 = \frac{ \sum_{\alpha}
	\left| \sum_n a_{3n}^{\alpha} e^{-b_n S'} \right|^2 }
{ \left( S' + \bar{S'} \right)
	\prod_{i \neq 3} \left( T_i + \bar{T}_i \right)
	\left| \eta \left( T_i \right) \right|^4 } \,.
\end{equation}

As pointed out by Brustein and Steinhardt \cite{Brustein} and Carlos
{\it et al} \cite{Carlos2}, the dilaton
provides the biggest obstacle to constructing a model of inflation
in superstrings.\footnote{At least if we don't assume something like
$S$-duality \cite{Font}, which allows some of the $b_n$ to be negative.}
Our model helps with the difficulty pointed out by Brustein and
Steinhardt, because $V_0$ can give $S'$ a suitable minimum during inflation
in much the same way as double gaugino condensation scenarios do in the
true vacuum \cite{Carlos}.
Since we are supposing that the $b_n$'s are positive, we need for this
purpose at least two distinct values of $n$ for some $\alpha$ so as to
obtain a minimum at a finite value of $S'$, and then at least one more term
with a different value of $\alpha$ to make $V_0$ nonzero.
A minimum with $V_0 > 0$ and mass greater than $H$ can then be obtained for
reasonable, but significantly constrained, values of the $a$'s and $b$'s.
There is, though, still the problem pointed out by Brustein and
Steinhardt that for
a potential of the form of $V_0$, and for generic
initial values, $S$ will tend to roll past the desired minimum and on
to the minimum at $S = \infty$. As this is also a problem for the
true vacuum it should be regarded as a problem for the assumption of all
positive $b_n$'s rather than of the model of inflation.
It might be solved by anthropic arguments, which in any case seem likely
to be needed because of the
huge degeneracy in the superstring vacuum.

There remains the problem that the $V_0$-induced
expectation value for $S'$ is likely to be different from its
vacuum expectation value after inflation because $V_0$ disappears at the
end of inflation. As pointed out by Carlos {\it et al}
\cite{Carlos2}, this might lead to cosmological problems because
$S'$ will in general be left far from its minimum at the end of inflation.
We do not address that difficulty here.

Next consider the moduli $T_i$ for $i \neq 3$.
The function $ ( T_i + \bar{T}_i ) | \eta ( T_i ) |^4 $ has its maxima at
$ T_i = e^{i\pi/6} $ and points equivalent under modular transformations, and
\begin{equation}
\left. ( T_i + \bar{T}_i ) | \eta ( T_i ) |^4
\right|_{T_i = e^{i\pi/6}+\sqrt{3}\, t_i}
= \sqrt{3} \left| \eta \left( e^{i\pi/6} \right) \right|^4
\left[ 1 - |t_i|^2 + O \left( t_i^3 \right) \right] \,,
\end{equation}
where $ \left| \eta \left( e^{i\pi/6} \right) \right| \simeq 0.8006 $.
Therefore $V_0$ is minimised for $ T_i = e^{i\pi/6} , i \neq 3$, and since,
to lowest order in $\delta^{GS}$ and the matter fields, the canonically
normalised $T_i$ fields (see Eq.~(\ref{ss3})) have masses $\sqrt{V_{0}}$
there, they will be firmly anchored during inflation. However, after
inflation they will be left sitting far from their true vacuum minima
(which are close to T=1.23 in some models \cite{Carlos}) potentially giving
cosmological problems \cite{Carlos2}. Note that if the $a_{3n}^{\alpha}$'s
were functions of the $T_i$ for $i \neq 3$, then they would merely shift the
$T_i$'s expectation values during inflation, whilst if they depended on $T_3$
they would fix $T_3$ during inflation, simplifying the following discussion.

Next, the canonically normalised $\phi_{i}^{\beta}$ matter fields
(see Eq.~(\ref{ss3})), for $i \neq 3$, acquire masses
$\sqrt{V_{0}} \simeq \sqrt{3}\,H$ (see Eq.~(\ref{ss7})), making them
unviable as inflatons and firmly fixing them during inflation.

Having argued for the stability of $V_0$ during inflation, let us consider
the possible inflatons.
The canonically normalised $\phi_{3}^{\beta}$ fields acquire a mass squared
(to lowest order in $\delta^{\rm GS}$ and the matter fields)
\begin{equation}
m_{\rm GS}^{2} =
\frac{\delta^{\rm GS}_{3} V_{0}}{4\pi^2 \langle S+\bar{S} \rangle} \,,
\end{equation}
much less than $V_0$ (assuming that we are in the perturbative regime
so that the loop corrections are small).
Note that here $\langle S+\bar{S} \rangle$ is the expectation value of the
dilaton during inflation, which is probably different from its value in our
vacuum. However, it may be reasonable to assume that it has a similar value
during inflation as now because, in both cases, we need to be in the
perturbative regime, Re$\,S \gsim 1$, but avoid the potentially runaway
behaviour at large Re$\,S$ \cite{Brustein}.
Also, $V_{\rm infl}$ is minimised for $ T_3 = e^{i\pi/6} $. Defining
$ T_3 = e^{i\pi/6} + \sqrt{3}\, t_3 $, and assuming
$\left| t_3 \right| \ll 1$ (and $\left| \phi_{3}^{\beta} \right| \ll 1$,
which we have been assuming all along), the $\phi_{3}^{\beta}$ and $T_3$
dependence of the inflationary potential energy is given by
\begin{equation}
V_{\rm infl} = V_{0} \left[ 1 +
\frac{\delta^{\rm GS}_{3}}{4\pi^2 \langle S+\bar{S} \rangle}
\left( \sum_{\beta} \left| \Phi_{3}^{\beta} \right|^2
+ \left| t_3 \right|^2 \right) + \ldots \right] \,,
\end{equation}
where $\Phi_{3}^{\beta}$ and $t_3$ are the canonically normalised
$\phi_{3}^{\beta}$ and $T_3$ fields. Then defining the canonically
normalised (inflaton) field $\phi = \sqrt{2}\, \sqrt{ \sum_{\beta}
\left| \Phi_{3}^{\beta} \right|^2 + \left| t_3 \right|^2 }$,
and making the reasonable assumption that the orthogonal degrees of
freedom are time independent, we get the potential during inflation
\begin{equation}
V_{\rm infl} = V_{0} \left[ 1 + \frac{1}{2}
\frac{\delta^{\rm GS}_{3}}{4\pi^2 \langle S+\bar{S} \rangle} \phi^2
\right] \,.
\end{equation}
Therefore, from Section~\ref{tilt}, the density perturbations produced
during inflation will have a spectral index
\begin{equation}
n = 1 + \frac{\delta^{\rm GS}_{3}}{2\pi^2 \langle S+\bar{S} \rangle} \,,
\end{equation}
directly related to fundamental superstring parameters. For example,
taking $\langle S+\bar{S} \rangle = 4$ \cite{Carlos} and
$\delta^{\rm GS}_{3} = 5$ \cite{GS} gives $n = 1.06$.

For $\delta^{\rm GS}_{3} = 0$ \cite{GS}, $T_3$ and $\phi_{3}^{\beta}$
receive no contribution to their potential from $V_{\rm infl}$ and so either
the terms in Eq.~(\ref{ss6}) neglected in Eq.~(\ref{ss65}) or the terms in
Eq.~(\ref{ss2}) neglected in Eq.~(\ref{ss6}) will dominate.
In the first case, if the $\partial W / \partial \phi_{i}^{\alpha}$ terms
for $i \neq 3$ are non-zero but still much smaller than the $i=3$ terms
then they could provide $T_3$ and the $\phi_{3}^{\beta}$ with masses
$\ll H$ in the same way as the $i=3$ terms provide the $T_i$ and
$\phi_{i}^{\beta}$ for $i \neq 3$ with masses of order $H$. Note that if
the $\partial W / \partial \phi_{i}^{\alpha}$ terms for $i \neq 3$ are of
the same order as the $i=3$ terms then all the fields will have masses of
order $H$ and inflation will not be possible.
In the latter case, the neglected terms are the terms that are thought to
provide the soft supersymmetry breaking terms in our vacuum and so, during
inflation, they might also provide the $\phi_{3}^{\beta}$ with soft
supersymmetry breaking mass terms and give $T_3$ a minimum with a
mass of the same order. However, as the expectation value of the dilaton
during inflation is likely to be different from its value in our vacuum,
the soft supersymmetry breaking scale is also likely to be different. Thus
for $\delta^{\rm GS}_{3} = 0$, the results will depend on the specific
superpotential.

Finally let us consider briefly the possibility that $\phi\gsim1$.
Then  Eq.~(\ref{ss7}) will no longer be a good approximation.
For example, consider the simplest case of constant $T_3$ and, without
loss of generality, one $\phi_{3}^{\beta}$ field which we will call
$\phi_3$. Then, to lowest order in $\delta^{\rm GS}$ and the other matter
fields, $\phi_3$'s kinetic term is
\begin{equation}
\frac{\partial^2 K}{ \partial \phi_3 \partial \bar{\phi}_3 }
\left| \partial_\mu \phi_3 \right|^2 =
\frac{ T_3 + \bar{T}_3 }{ \left( T_3 + \bar{T}_3 - \left| \phi_3 \right|^2
	\right)^2 } \left| \partial_\mu \phi_3 \right|^2 \,.
\end{equation}
Therefore the canonically normalised real field $\phi$ corresponding to
$ \left| \phi_3 \right| $ is given by
\begin{equation}
\left| \phi_3 \right| =
\sqrt{ T_3 + \bar{T}_3 } \, \tanh \frac{\phi}{\sqrt{2}} \,.
\end{equation}
Note that $\phi \gsim 1$ corresponds to $|\phi_3| \sim
\sqrt{ T_3 + \bar{T}_3 }$ but still $|\phi_3| < \sqrt{ T_3 + \bar{T}_3 }$.
This is the only place where we will relax the assumption of
$|\phi_3| \ll \sqrt{ T_3 + \bar{T}_3 }$. One of the problems of relaxing
this assumption is that we neglected the terms in Eq.~(\ref{ss2})
proportional to $1/X_3 = \cosh^2 (\phi/\sqrt{2}) / ( T_3 + \bar{T}_3 )$.
\footnote{Note that these terms, although not inflationary, could lead to
a scaling $a \propto t$ of the scale factor of the universe, taking us down
from the Planck scale to the energy scale at which inflation proper starts.}
This will be reasonable for $\phi \ll 1$ but is unlikely to be so for
$\phi \gg 1$. However it may just be acceptable for $\phi \sim 1$,
and so, making the big assumption that the derivation of Eq.~(\ref{ss65})
is still valid for $ \phi \gsim 1$, we get the inflationary potential
\begin{equation}
V_{\rm infl} = V_0 \left[ 1
+ 2 \frac{\delta^{\rm GS}_{3}}{4\pi^2 \langle S+\bar{S} \rangle}
\ln \cosh \frac{\phi}{\sqrt{2}} \right] \,.
\end{equation}
For example, taking $\langle S+\bar{S} \rangle = 4$ \cite{Carlos},
$\delta^{\rm GS}_{3} = 15$ \cite{GS} and $\phi_{\rm inst} \sim V_{0}^{1/4}$
could give an observable signature of superstrings in the varying spectral
index of the density perturbations produced during inflation.
Note that we must be in the vacuum dominated regime for the loop expansion
(expansion in $\delta^{\rm GS} / 4\pi^2 \langle S+\bar{S} \rangle$)
to be reliable.

\subsubsection*{A specific model}

The arguments that we have given suggest that
false vacuum inflation can be achieved, provided
that the superpotential satisfies certain conditions. We have not however
demonstrated that such a superpotential exists, nor have we discussed the
instability mechanism which ends inflation (see Sections~2 and~3).
We therefore end by showing how things work out with a specific choice for
the perturbative part of the superpotential. The form we choose is
\begin{eqnarray}
W_{\rm pert} & = &
\left( \hat{\phi}_{1}^{(1)} \hat{\phi}_{2}^{(1)}
+ \hat{\phi}_{1}^{(2)} \hat{\phi}_{2}^{(2)}
+ \hat{\phi}_{1}^{(3)} \hat{\phi}_{2}^{(3)} \right) \hat{\phi}_{3}
+ \left( \hat{\phi}_{1}^{(1)} \hat{\phi}_{2}^{(4)}
+ \hat{\phi}_{1}^{(4)} \hat{\phi}_{2}^{(1)} \right) \tilde{\phi}_{3}
\nonumber \\
 & & + \left( \hat{\phi} \rightarrow \check{\phi} \right) \,,
\end{eqnarray}
where $\hat{\;\;}$, $\check{\;\;}$, $\tilde{\;\;}$ and ${\;}^{(\alpha)}$
correspond to the generic ${\;}^{\alpha}$ label used previously.
We also assume that $\hat{\phi}_{i}^{(\alpha)}$ and
$\check{\phi}_{i}^{(\alpha)}$ for $i=1,2$ and $\alpha=2,3$
acquire the expectation values
\begin{equation}
\label{ss8}
\hat{\phi}_{i}^{(\alpha)} = \hat{\Lambda}_{i}^{(\alpha)} \left( S,T \right)
= \frac{ \hat{a}_{i}^{(\alpha)} e^{-\hat{b}_{i}^{(\alpha)} S'} }
{ \left[\eta \left(T_i \right) \right]^2 } \,,
\end{equation}
and similarly for $\check{\phi}$. The form of these expectation values
is motivated by gaugino condensation scenarios \cite{Carlos}.
These expectation values might be induced by the D-term part of the
scalar potential or, more directly, by a nonperturbative part of the
superpotential. We will assume $\tilde{\phi}_{3}$ is D-flat because
it will become (part of) the inflaton.

Substituting into Eq.~(\ref{ss6}) we get
\begin{eqnarray}
V & = & \frac{1}{ X_2 X_3
	\left( Y + \frac{1}{4\pi^2}\delta^{\rm GS}_{1} \right) }
\left[ \left| \hat{\phi}_{2}^{(1)} \hat{\phi}_{3}
	+ \hat{\phi}_{2}^{(4)} \tilde{\phi}_{3} \right|^2
+ \left| \hat{\Lambda}_{2}^{(2)} \hat{\phi}_{3} \right|^2
+ \left| \hat{\Lambda}_{2}^{(3)} \hat{\phi}_{3} \right|^2
+ \left| \hat{\phi}_{2}^{(1)} \tilde{\phi}_{3} \right|^2
+ \left( \hat{\;} \rightarrow \check{\;} \right) \right] \nonumber \\
 & & + \left( {\rm subscript} \; 1 \leftrightarrow {\rm subscript} \; 2
\right) \nonumber \\
 & & + \frac{1}{ X_1 X_2
	\left( Y + \frac{1}{4\pi^2}\delta^{\rm GS}_{3} \right) }
\left[ \left| \hat{\phi}_{1}^{(1)} \hat{\phi}_{2}^{(1)}
	+ \hat{\Lambda}_{1}^{(2)} \hat{\Lambda}_{2}^{(2)}
	+ \hat{\Lambda}_{1}^{(3)} \hat{\Lambda}_{2}^{(3)} \right|^2
+ \left( \hat{\;} \rightarrow \check{\;} \right)
\right. \nonumber \\
 & & \left.
+ \left| \hat{\phi}_{1}^{(1)} \hat{\phi}_{2}^{(4)}
	+ \hat{\phi}_{1}^{(4)} \hat{\phi}_{2}^{(1)}
	+ \left( \hat{\phi} \rightarrow \check{\phi} \right) \right|^2
\right] \,.
\end{eqnarray}
For $|\tilde{\phi}_{3}| > 0$ this potential is minimised for
$ \hat{\phi}_{3} = \check{\phi}_{3}
= \hat{\phi}_{1}^{(4)} = \check{\phi}_{1}^{(4)}
= \hat{\phi}_{2}^{(4)} = \check{\phi}_{2}^{(4)} = 0 $.
Then
\begin{eqnarray}
\label{ss9}
V & = &
\frac{ \left( \left| \hat{\phi}_{2}^{(1)} \right|^2
	+ \left| \check{\phi}_{2}^{(1)} \right|^2 \right)
	\left| \tilde{\phi}_3 \right|^2 }
{ X_2 X_3 \left( Y + \frac{1}{4\pi^2}\delta^{\rm GS}_{1} \right) }
+ \left( {\rm subscript} \; 1 \leftrightarrow {\rm subscript} \; 2 \right)
\nonumber \\
 & &
+ \frac{ \left| \hat{\phi}_{1}^{(1)} \hat{\phi}_{2}^{(1)}
	+ \hat{\Lambda}_{1}^{(2)} \hat{\Lambda}_{2}^{(2)}
	+ \hat{\Lambda}_{1}^{(3)} \hat{\Lambda}_{2}^{(3)} \right|^2
	+ \left( \hat{\;} \rightarrow \check{\;} \right) }
{ X_1 X_2 \left( Y + \frac{1}{4\pi^2}\delta^{\rm GS}_{3} \right) } \,.
\end{eqnarray}
Defining the canonically normalised fields
$\tilde{\Phi} \propto \tilde{\phi}_{3}$,
$\hat{\Psi}_1 \propto \hat{\phi}_{1}^{(1)}$,
$\hat{\Psi}_2 \propto \hat{\phi}_{2}^{(1)}$,
$\hat{\Lambda}^2 \propto \hat{\Lambda}_{1}^{(2)} \hat{\Lambda}_{2}^{(2)}
+ \hat{\Lambda}_{1}^{(3)} \hat{\Lambda}_{2}^{(3)}$,
and similarly for the checked symbols, then working to lowest order in
$\delta^{\rm GS}$ and the matter fields, we obtain
\begin{equation}
V = \frac{1}{ S + \bar{S} }
\left[ \left| \hat{\Psi}_1 \hat{\Psi}_2 + \hat{\Lambda}^2 \right|^2
+ \left( \left| \hat{\Psi}_1 \right|^2 + \left| \hat{\Psi}_2 \right|^2
	\right) \left| \tilde{\Phi} \right|^2
+ \left( \hat{\;} \rightarrow \check{\;} \right) \right] \,.
\end{equation}
Proceeding as in Section~\ref{sa} then gives
\begin{equation}
\label{ss10}
V = \frac{1}{ S + \bar{S} }
\left[ \frac{1}{16} \left(\hat{\psi}^2 - 4|\hat{\Lambda}|^2 \right)^2
+ \frac{1}{4} \tilde{\phi}^2 \hat{\psi}^2
+ \left( \hat{\;} \rightarrow \check{\;} \right) \right] \,.
\end{equation}
For $\tilde{\phi} > \sqrt{2}\, \max\{ \hat{\Lambda} , \check{\Lambda} \}$,
this is minimised for $\hat{\psi}=\check{\psi}=0$, giving the false vacuum
energy density
\begin{equation}
V = V_0 = \frac{ |\hat{\Lambda}|^4 + |\check{\Lambda}|^4 }{ S + \bar{S} } \,.
\end{equation}
as discussed above. The higher order terms give the inflaton a potential,
also as discussed above. However, at $\tilde{\phi} = \tilde{\phi}_{\rm inst}
= \sqrt{2}\, \max\{ \hat{\Lambda} , \check{\Lambda} \}$ an instability
sets in ending inflation in a manner similar to that described in
Sections~2 and~3.

\section{A First-Order Model}
\label{MODELS}
\setcounter{equation}{0}
\def\theequation{\thesection.\arabic{equation}}

The one context in which the dynamical effect of more than one scalar field
during inflation has been considered in some detail in the literature is in
models of inflation ended by a first-order phase transition, where a field
must tunnel from the metastable false vacuum, through a classically forbidden
region, to the true vacuum. In the case of a single scalar field (Guth's old
inflation model \cite{Guth}) the metric rapidly reaches the static de Sitter
metric with a fixed nucleation rate to the true vacuum and the transition must
either complete at once (without sufficient inflation) or not at all. The
critical parameter here is the percolation parameter, $p$, the average number
of bubbles nucleated per Hubble volume per Hubble time. To complete the
transition $p$ must exceed some critical value, $p_{\rm cr} = {\cal O}(1)$
\cite{GuthWeinberg}. By introducing a second scalar field which can evolve
with time, $p$ can grow allowing sufficient inflation before the transition
completes.

To incorporate such a first-order transition into our model we must extend our
basic potential, Eq.~(\ref{FULLPOT}), to include asymmetric terms which can
break the degeneracy of the two vacuum states at low energies. Thus we will
consider the more general potential,
\begin{equation}
\label{ASYMPOT}
V(\phi,\psi) = \frac{1}{4}\lambda \left( M^4 + \psi^4 \right)
	+ \frac{1}{2} \alpha M^2 \psi^2
	- \frac{1}{3} \gamma M \psi^3
	+ \frac{1}{2} m^2 \phi^2 + \frac{1}{2} \lambda' \phi^2 \psi^2 \,.
\end{equation}
The cubic term spoils the degeneracy, and choosing $\alpha$ greater or less
than zero determines whether $\psi=0$, $\phi=0$ is a local minimum or saddle
point respectively. Thus in addition to the two mass scales $M$ and $m$ we now
have four dimensionless coupling constants $\lambda$, $\lambda'$, $\alpha$ and
$\gamma$. Requiring the energy density of the true vacuum to be zero
($V(0,\psi_{\rm true})=0$) can be used to specify $\gamma$, say, in terms of
$\alpha$ and $\lambda$. Thus we have one more free parameter, $\alpha$, than
in our second-order model (which corresponds to the particular case
$\alpha=-\lambda$, $\gamma=0$).

At large values of $|\phi|$
(where $\lambda'(\phi/M)^2>(\gamma^2/4\lambda)-\alpha$) the potential has
only one
turning point with respect to $\psi$, a minimum at $\psi=0$, while for smaller
values of $|\phi|$ a second minimum appears, initially as a point of inflection
at
$\psi=\gamma/2\lambda$. Although it is this second minimum that develops into
the true vacuum with $V=0$ when $\phi=0$, for $\gamma\neq0$ it initially
has an energy density greater than that of the false vacuum
so that if the fields follow the ``path of least resistance'' they
will remain in the false vacuum for $\alpha+\lambda'(\phi/M)^2>0$.

\subsection{Inflationary dynamics}

While $\psi$ is restricted to the false vacuum ($\psi=0$) the potential for
$\phi$ remains that given in Eq.~(\ref{PHIPOT}), and the dynamics are the same
as considered in Section \ref{INFL}, except that the effective mass of the
$\psi$ field is now
\begin{equation}
M_{\psi}^2 = \alpha M^2 + \lambda' \phi^2  \equiv \tilde{\alpha}(\phi) M^2
	\,,
\end{equation}
and a second-order transition is not possible for $\alpha>0$. Instead the
transition must proceed by nucleating bubbles of the true vacuum and the
end-point $\phi_{\rm inst}$ is replaced by the critical value $\phi_{\rm cr}$
where the percolation parameter reaches $p_{\rm cr}$.

Notice then that inflation ends at
\begin{equation}
\phi_{\rm end} = {\rm max} \{ \phi_\epsilon,\phi_{\rm cr} \} \,,
\end{equation}
where the slow-roll condition may break down at $\phi_\epsilon$ (defined as in
Section \ref{INFL}) before the true vacuum percolates. If this is the case
then we are again in the inflaton dominated limit, the false vacuum energy
density is negligible ($\lambda M^4 \ll m^2\phi_{60}^2$) and the constraints
are exactly the same as when the eventual transition to true vacuum is
second-order.  The precise mechanism of the phase transition becomes
irrelevant as this now occurs after inflation has ended.
After passing $\phi_\epsilon$ the field reaches $\phi=0$ within one Hubble
time.  Unlike the second-order model this does not immediately cause an
instability, and oscillations about $\phi=0$ could be sufficiently damped to
restart inflation if $\phi_{\rm cr}$ lies very close to zero.  However even in
this case we can show that the number of $e$-foldings, given by
Eq.~(\ref{EFOLD}), during any subsequent stage of inflation
\begin{equation}
N < \frac{1}{8} + \frac{1}{4} \ln \left( \sqrt{
	\frac{\lambda'}{16\pi\lambda}} \, \frac{\mpl}{M} \right) \,,
\end{equation}
must be very small.

Thus we will consider only the vacuum dominated branch in what follows, where
we may take $\lambda M^4 \gg m^2\phi_{\rm cr}^2$. Another reason for doing
this is that we will have to ignore the evolution of the $\phi$ field while
calculating the nucleation rate for $\psi$ from the false to the true vacuum.
The correct two--field result is not known so, in common with all other models
of first--order inflation, we will calculate instead the tunnelling rate for
the
quasi--static potential $V(\psi)$.  We would only expect this to be valid if
$m\lsim H$, which is indeed guaranteed if we are in the vacuum dominated
regime.

The percolation parameter is then given by
\begin{equation}
p \simeq \frac{\lambda M^4}{4 H^4} \exp ( -S_E ) \,,
\end{equation}
where the term in the exponential is the Euclidean action of the tunnelling
configuration \cite{Coleman}, recently given for first-order quartic
potentials $V(\psi)$ by Adams \cite{Adams} as $S_E=2\pi^2B_4/\lambda$, with
$B_4$ a numerically calculated monotonically increasing fitting function of
the parameter
\begin{equation}
\label{DELTA}
\delta (\phi) \equiv \frac{9\lambda\tilde{\alpha}}{\gamma^2} \,.
\end{equation}

In our model $\delta(\phi)$ decreases as $\phi$ rolls down its potential
during inflation, until $S_E$ is sufficiently small for the percolation
parameter to reach unity allowing the first-order transition to complete. This
corresponds to
\begin{equation}
S_{\rm cr} = \ln \frac{\lambda M^4}{4 p_{\rm cr} H^4} \simeq 4 \ln
	\frac{\mpl}{M} \,.
\end{equation}
Figure 3 shows the corresponding value of $\delta_{\rm cr}$ required for the
transition to complete at different values of the false vacuum energy density.
Clearly for a given $\alpha$ there is a lower bound,
$\delta>\delta_0=9\lambda\alpha/\gamma^2$, and a corresponding bound on the
nucleation rate, so the first-order transition cannot complete when the energy
density of the false vacuum is above a given value.  For instance, if
$\lambda=1$, and $\alpha \gsim 1$ then the transition will never complete for
$M\gsim10^{14}$GeV.

Assuming then that $\alpha$ is sufficiently small (i.~e.~$\delta_0<\delta_{\rm
cr}(M)$), bubbles of the true vacuum will percolate at $\phi_{\rm
cr}=\sqrt{\lambda/\lambda_{\rm eff}'} \, M$ , where, to utilise the results of
Section \ref{INFL}, we write
\begin{equation}
\lambda_{\rm eff}' = \frac{9\lambda^2}{\gamma^2(\delta_{\rm cr}-\delta_0)}
	\lambda' \,.
\end{equation}
To complicate the matter somewhat, $\lambda_{\rm eff}'$ is now a function of
$M$ through the dependence of $\delta_{\rm cr}$ on the energy density, but
this dependence is very weak for the energy scales significantly below the
Planck scale which we are interested in.  With this proviso then, the results
for the vacuum dominated branch of Section \ref{INFL} in the second-order
model may be carried through to the first-order model by replacing $\lambda'$
by $\lambda_{\rm eff}'$.

\subsection{Big bubble constraints}

The production of large true vacuum voids, nucleated early on during inflation
and swept up to astrophysical sizes by the subsequent expansion, can severely
constrain some models of first-order inflation \cite{Weinberg,LW1}. The
isotropy of the microwave background can be used to rule out the possibility
that there are any voids with a comoving size greater than about $20h^{-1}$Mpc
on the last scattering surface \cite{LW1}, which corresponds to a filling
fraction of less than about $10^{-5}$ for bubbles nucleated around $55$
$e$-foldings before the end of inflation.  This means that the percolation
parameter at this point during inflation must be less than $10^{-5}$,
requiring
\begin{equation}
S_{55} \gsim 4 \ln \frac{\mpl}{M} + 11.5 \,.
\end{equation}
This gives the second line in Figure 3 showing the minimum permissible value
of $\delta$ (denoted by $\delta_*$) at $55$ $e$-foldings before the end of
inflation at different false vacuum energy densities.

Obeying this extra constraint, $\delta_{55}\gsim\delta_*$, requires
$\tilde{\alpha}$ to be greater than a minimum value at this point and thus
\begin{equation}
\phi_{55} \gsim \gamma \sqrt{\frac{\delta_* - \delta_0}{9\lambda\lambda'}} M
	\,.
\label{PHI55}
\end{equation}
In the vacuum dominated regime the value of $\phi$ can be given as a function
of the number of $e$-foldings before the end of inflation from
Eq.~(\ref{EFOLD})
\begin{equation}
\phi \simeq \phi_{\rm cr} \exp \left( \frac{Nm^2 \mpl^2}{2\pi\lambda M^4}
 \right) \, ,
\end{equation}
which gives the constraint in Eq.~(\ref{PHI55}) as a constraint on the mass
scales
\begin{equation}
\frac{m^2 \mpl^2}{M^4} \gsim \frac{\pi\lambda}{55} \ln \left(
	\frac{\delta_* - \delta_0}{\delta_{\rm cr} - \delta_0} \right) \,.
\end{equation}
In other words, the mass of the $\phi$ field, $m$, must be large enough for
the decrease in the effective mass of the $\psi$ field during the last $55$
$e$-foldings of inflation to raise the percolation parameter from $10^{-5}$ to
unity. The numerical factor on the right-hand side of this equation is fairly
small, typically about $10^{-2}$ for $\lambda\sim1$, so this does not threaten
to force us out of the small $m$ limit. Clearly it is minimised for small
$\alpha$, as $\delta_0\to0$, but can become large if $\delta_{\rm cr}$ is too
close to $\delta_0$.

Normalising the parameters of the model by the observed density perturbations,
as described in Section \ref{INFL}, gives another relation between $m$ and
$M$ which, combined with the big-bubble constraint, provides limits on either
$m$ or $M$ alone:
\begin{eqnarray}
\label{BBBIGM}
\frac{M}{\mpl} & \gsim & (3\times10^{-5}) \frac{\pi\gamma}{55\lambda}
	\sqrt{\frac{\delta_{\rm cr}-\delta_0}{9\lambda'}} \ln \left(
	\frac{\delta_* - \delta_0}{\delta_{\rm cr} - \delta_0} \right) \,,
	\\
\label{BBSMALLM}
\frac{m}{\mpl} & \gsim & (3\times10^{-5})^2 \frac{\gamma^2}{\lambda^{3/2}}
	\left( \frac{\delta_{\rm cr}-\delta_0}{9\lambda'} \right)
	\left( \frac{\pi}{55} \right)^{5/2} \left( \ln
	\frac{\delta_*-\delta_0}{\delta_{\rm cr}-\delta_0} \right)^{5/2} \,.
\end{eqnarray}
For reasonable values of the coupling parameters, of order unity, we would
expect the right-hand side of Eq.~(\ref{BBBIGM}) to be $\sim 10^{-6}$ placing
a lower limit on $M$ of around $10^{13}$GeV in a first-order model, unless we
have $\delta_{\rm cr}$ very close to $\delta_0$. The other way to allow
first-order models at lower energy scales would be to introduce a strong
coupling $\lambda'$, much larger than unity, between the two fields, which is
clearly always possible as this enables only a small change in $\phi$ to
effect a large change in the bubble nucleation rate.

\subsection{Other first-order models}

The ability of a second scalar field to allow a first-order inflationary phase
transition to complete was first emphasised by La and Steinhardt
\cite{LaSteinhardt}. This is the basis of models of extended inflation based
on extensions to the gravitational lagrangian beyond the Einstein-Hilbert
action of general relativity \cite{LaSteinhardt,extinf,Amendola}.  In
Brans-Dicke gravity, for instance, the Ricci scalar appears in the action
coupled to a scalar field rather than Newton's constant and it is this growing
Brans-Dicke field, $\Phi \equiv \mpl^2$, which triggers the completion of
the phase transition in the $\psi$ field. However Linde \cite{Linde90} and
Adams and Freese \cite{AdamsFreese}, pointed out that this basic scenario can
also be realised in general relativity by coupling the inflaton to a second
scalar field. Linde used the same basic first-order potential $V(\psi)$ as we
have, although he used a Coleman-Weinberg type potential for $\phi$ rolling
down from $+\infty$ introducing a minimum at a non-zero value.  This would
have to be included in the minimum value of $\tilde{\alpha}$ and thus
$\delta$. Adams and Freese considered a specific interaction rather different
to ours where as $\phi$ rolled down its potential, the energy of the
false vacuum
state actually increased relative to the true vacuum, but their more general
discussion was clearly intended to include models such as the one we have
examined here.

The bubble nucleation constraints in terms of $\delta_{\rm cr}$ and
$\delta_{55}$ are independent of the type of first--order inflation being
considered.  Extended inflation models consider a first--order potential for
the inflaton which does not change during inflation. Thus $\delta$
remains a constant, as does the false vacuum energy density.  The time--varying
quantity here is the Planck mass which grows during inflation. Thus extended
inflation models proceed horizontally, from right to left across the parameter
space in Figure 3, completing the phase transition when $\delta_{\rm
cr}(M/\mpl)=\delta$. The general relativistic models considered here, and
those considered by Linde and by Adams and Freese, proceed almost vertically
as the false vacuum energy density remains approximately constant as $\delta$
decreases with $\phi$.  Because the percolation parameter $p$ is exponentially
dependent on the Euclidean action, $S_E$, it is relatively easy to evade the
big--bubble constraints in the general relativistic models varying $\delta$. In
models where only $M$ or $\mpl$ varies, the percolation parameter tends to
grow comparatively slowly making the big--bubble contraint much more severe,
especially as $V/\mpl^4$ and $V'/V$ are already constrained by density
pertubations at $60$ e-foldings \cite{LL1}.

This has led other authors
\cite{LIN2SC2,Laycock,Amendola} recently to consider models of extended
inflation
where non-minimal coupling can also change the shape of the `effective
potential', making $M_{\psi}^2=\xi R-\lambda M^2$ for instance.
In such cases inflation could
again end by a first-- or second--order transition. In a de Sitter metric the
Ricci scalar $R$ is a constant ($R=12H^2$) so a false vacuum dominated
universe in general relativity does not yield a time-varying mass. But in
Brans-Dicke gravity for instance, where the dominant coupling to the Ricci
scalar is via the Brans-Dicke field (rather than a constant) the expansion is
power-law \cite{Nariai} rather than exponential and $R\propto t^{-2}$,
triggering an instability when $R\leq\sqrt{\lambda/\xi} \, m$. Similar models
have been proposed in higher order gravity theories, coupling the $\psi$ field
to $R^2$ terms \cite{Amendola}. These models extending the gravity lagrangian
can be re-written in terms of a general relativistic model with two
interacting scalar fields (the defect field and a dilaton field that acts as
the inflaton) using a conformally rescaled metric \cite{Maeda}. But the scalar
field lagrangian in this case is rather different from our model as not only
$M_{\psi}$ but all the mass scales are changed by the dilaton field. These
first--order models thus correspond to a more complicated path on Figure 3,
and by making $\delta$ a function of time can also evade the big--bubble
constraint.

\section{Discussion and Conclusions}
\label{CONC}
\setcounter{equation}{0}
\def\theequation{\thesection.\arabic{equation}}

In conclusion, models of inflation based on Einstein gravity, but driven by
a false vacuum, offer a range of new possibilities for both theory and
phenomenology.

On the particle physics side, we have shown how false vacuum
inflation points to new possibiities for model building.
 In particular, we have shown
that it can occur in a class of supergravity models implied by orbifold
compactification of superstrings.
One outcome of that discussion was the intriguing
possibility of obtaining a handle on the superstring orbifold,
through the fact that
one-loop corrections might be the dominant effect determining
the spectral index.
Much remains to be done of course.
For instance, although we have
exhibited a toy model for the scalar field
sector of the string derived supergravity theory,
we have made no attempt to put it in the context of
a realistic model involving other fields as well.
In particular we have not tried to extend to supergravity
the identification of the false vacuum with
that of Peccei-Quinn symmetry, which we found was both viable and
attractive in the context of global supersymmetry.

In terms of direct cosmological phenomenology,
false vacuum dominated inflation
offers the unusual
option of a spectral index for the density perturbations exceeding unity,
though we have demonstrated that with the COBE normalisation the deviation can
only be rather modest with a plausible maximum of around $n = 1.14$. There is
however additional interest in that one expects topological defects to form as
the false vacuum decays; because essentially all the energy density is
available to go into the defect fields, the energy available is much greater
than in usual models where reheating is required first, redistributing the
energy into a large number of fields. Because of this, structure-forming
defects are comfortably compatible with our inflation model when the masses
are towards the top of their allowed ranges.

We have also made a preliminary investigation of the details of the phase
transition in different regimes, though much remains to be done. For a
second-order phase transition, results already exist in the literature
describing the inflaton dominated regime. We have demonstrated that, barring
very weak couplings, the phase transition proceeds very rapidly in the vacuum
dominated regime, but have been unable to develop a solid understanding of the
statistics of the defects produced in such a transition. In the first-order
case, where the transition completes via bubble nucleation, we have gone on to
calculate the bubble distribution and the constraints upon it. We note that
first-order inflation models based on Einstein gravity are generally easier to
implement than those of the extended inflation type.

That one can have both structure-forming topological defects and inflation
raises a host of possible structure formation scenarios, as one could choose
to utilise only one of these two or a combination of the two. It is believed
\cite{Albert} that for a given size of density perturbation
(i.~e.~perturbation in the gravitational potential), defects give a larger
microwave background temperature anisotropy, by a factor of a few. One could
therefore arrange for defects to be the source of a component of the COBE
signal while having only a modest effect on structure formation; alternatively
one could aim to have inflation and defects contributing roughly equally to
structure formation in which case the defects would be predominant in the
microwave background. It is conceptually (and calculationally) preferable to
take the option of using only one source, lowering the energy scale of the
other to make its effects negligible, but one should be aware that the
required scales of the two are similar, and should a realistic model along our
suggested lines be devised it would not be a particular surprise should both
contributions have a role to play.

%%%%%%%%%%%%%%%%%%%%%%%%%%%%%%%%%%%%%%%%%%%%%%%%%%%%%%%%%%%%%%%%%%%%%%

\section*{Acknowledgements}

EJC and DW are supported by the SERC, ARL by the SERC and the Royal Society
and EDS by a JSPS Postdoctoral Fellowship and Monbusho Grant-in-Aid for
Encouragement of Young Scientists, No.\ 92062. EJC and ARL acknowledge the
hospitality of the Aspen Center for Physics, ARL that of Fermilab and
and DHL that of CERN; part of this work was carried out at each of these
places.
We would like to thank Mark Hindmarsh and Andrei Linde for helpful
discussions, and Kiwoon Choi, Burt Ovrut and Graham Ross for helpful comments
on
early drafts of the paper.
ARL and DW acknowledge the use of the STARLINK computer system at
the University of Sussex.
%%%%%%%%%%%%%%%%%%%%%%%%%%%%%%%%%%%%%%%%%%%%%%%%%%%%%%%%%%%%%%%%%%%%%%
\frenchspacing
%%%%%%%%%%%%%%%%%%%%%%%%%%%%%%%%%%%%%%%%%%%%%%%%%%%%%%%%%%%%%%%%%%%%%%

%%%%%%%%%%%%%%%%%%%%%%%%%%%%%%%%%%%%%%%%%%%%%%%%%%%%%%%%%%%%%%%%%%%%%%
\section*{Figure Captions}

\vspace{24pt}
\noindent
{\em Figure 1}\\
The solid line shows the locus of $M$ and $m$ which satisfy the COBE
normalisation for $\lambda = \lambda' = 1$. The two analytic branches are
clearly seen. The dot-dashed line indicates the analytic solution for the
extreme vacuum dominated branch, as utilised by Linde \cite{LIN2SC2}; the
deviation of the exact solution from it is caused by the increasing
significance of the exponential term in Eq.~(\ref{cmb}) which is included in
the parametric analytic solution Eq.~(\ref{XXXX}), as used in \cite{MML} and
indicated here by the dotted line. The dotted line (hidden under the solid for
most of its length) terminates when the regime becomes invalid, though it
actually extends somewhat beyond the exact solution because we have
interpreted $\ll$ as $\lsim$ in places.

\vspace{24pt}
\noindent
{\em Figure 2}\\
The spectral index $n$ is shown for the exact COBE normalised models of Figure
1.

\vspace{24pt}
\noindent
{\em Figure 3}\\
$\delta_{cr}$ and $\delta_*$ plotted as functions of $V^{1/4}/\mpl$ for
first-order inflation. 55 e-foldings from the end of inflation
$\delta_{55}$ must lie above the dotted line, $\delta_*$, but then reach
the solid line, $\delta_{cr}$, to bring inflation to an end. The two
trajectories, plotted as dashed lines, represent the typical evolution of
$\delta$ and $M/m_{\rm PL}$ for (a) extended inflation where
$\delta=$constant, and (b) false vacuum inflation in Einstein gravity
where $V^{1/4}/\mpl\simeq$ constant.

%\vspace{24pt}
%\noindent
%{\em Figure 3}\\
%CAPTION
%%%%%%%%%%%%%%%%%%%%%%%%%%%%%%%%%%%%%%%%%%%%%%%%%%%%%%%%%%%%%%%%%%%%%%
\end{document}